\documentclass[preprint]{nrc1}

\usepackage{amsmath, amssymb, bm, cite, graphicx, graphics, color,mathrsfs}

\newcommand{\Lie}[0]{{\cal L}\, }

\newcommand{\tl}{\theta_{(\ell)}}
\newcommand{\tn}{\theta_{(n)}}
\newcommand{\scri}{\mathscr{I}}

\newcommand{\nn}{\nonumber}
\newcommand{\be}{\begin{equation}}
\newcommand{\ee}{\end{equation}}
\newcommand{\bea}{\begin{eqnarray}}
\newcommand{\eea}{\end{eqnarray}}

\newcommand{\tq}{\tilde{q}}
\newcommand{\hu}{\hat{u}}

\newcommand{\htau}{\hat{\tau}}
\newcommand{\hr}{\hat{r}}

\newcommand{\tom}{\tilde{\omega}}

\newcommand{\vH}{{}^{{}^{(H)}} \! \! \mbox{\boldmath$\epsilon$}}

\newcommand{\vS}{\widetilde{\mbox{\boldmath$\epsilon$}}}

\newcommand{\bdv}{\mathbf{d}\mbox{\boldmath{$v$}}}

\newcommand{\cV}{\mathcal{V}}

\journal{ }

\begin{document}

\title{Black hole boundaries}
\author {Ivan Booth}
\address{Department of Mathematics and Statistics, Memorial
 University\\
St. John's, Newfoundland and Labrador, A1C 5S7, Canada \\
ibooth@math.mun.ca}
\shortauthor{I. Booth}

\maketitle

\begin{abstract}
Classical black holes and event horizons are highly non-local objects, defined in relation to  the causal
past of future null infinity. Alternative, quasilocal characterizations of black holes 
are often used in mathematical, quantum, and numerical relativity. 
These include apparent, killing, trapping, isolated, dynamical, and slowly evolving  horizons. All of these
are closely associated with two-surfaces of zero outward null expansion. This paper reviews the traditional definition
 of black holes and provides an overview of some of the more recent work on alternative horizons. 
\end{abstract}

\section{Introduction}

Black holes have played a central role in gravitational physics for over 40 years. Early on, they were largely 
creatures of mathematical relativity and it was in the 1960s and 1970s that many of their classical properties were
investigated. These included both detailed studies of the geometry and physics of stationary exact solutions as well 
as proofs of general results such as the singularity, uniqueness, and area increase theorems (for reviews, 
see the standard texts such as \cite{hawkellis, chandra, wald}). Since then the 
mathematical interest has continued but they have also found wide application in other areas of 
gravitational research. Notably, the discovery in the early 70s of the laws of black hole mechanics
\cite{bch} and subsequent identification of those laws with the laws of thermodynamics \cite{bekhawk} has
reverberated through studies of quantum gravity ever since. They have also found a key place in gravitational
astrophysics and the observational evidence for their existence is very strong and getting better all 
the time \cite{narayan}. Theoretically, they are 
closely linked to some of the most powerful processes known to science including both supernovae
and active galactic nuclei \cite{krolik}. Physical processes involving black holes (such as binary inspirals and 
gravitational collapse) are widely expected to be some of the brightest sources of gravitational waves and based
on these expectations, gravitational wave observatories have been built all over the world \cite{ligo}.


All of that said, the standard mathematical definition of a black hole \cite{hawkellis} as laid down over 30 years ago is largely disconnected from these applications. 
The extent of a black hole is not directly defined by strong gravitational fields 
(equivalently large spacetime curvatures) or any other 
locally defined quantities. 
Instead, it is a global property of the causal structure of an entire spacetime; one cannot properly identify 
a black hole region without that global knowledge. We will discuss the rigourous definition in more detail 
later in this review, but for now paraphrase it in the usual way by saying that a black hole
is a region of spacetime from which no signal can ever escape. As a definition this is intuitively satisfying
but complications quickly arise once we try to properly define ``ever" and ``escape". To do this in an invariant and mathematically elegant way we are inevitably lead to a consideration of the structure of 
spacetime at infinity --- only after waiting an infinite amount of time can one really say whether or not
 a signal will get out and only once it is an infinite distance from the hole can we really say that it has 
escaped. 

Thus, this definition depends on observations which can only be made ``at infinity" and so the
exact extent of a black hole can only be determined by omniscient observers. In practice, of course, such 
observations are impossible and so, as a physicist, one might adopt a more practical 
viewpoint of deciding on the existence/extent of a black hole by making measurements at large distances after waiting for long periods of time. Even then though the definition remains teleological and non-local and 
so philosophically problematic. For example we will see that an event horizon will always expand in 
area until the last bit of matter/energy that will ever pass through it does so. Because it is identified
in the far future, its evolution reflects that knowledge. 


Given these difficulties, there has always been some discussion of other, (quasi)local, 
characterizations of black holes. Instead of the causal structure, these alternative definitions are geometric 
and identify black holes\footnote{``Black hole" is used in a general sense here as the kind of object that, for example, forms
during a gravitational collapse. Henceforth we will often use it in this common way, and generally add ``causally defined", ``standard" 
or some other identifying phrase in situations when we instead mean the specific classical definition.}
with regions of strong gravitational fields and in particular
with trapped surfaces : closed and spacelike two-surfaces which have the property that all null geodesics which intersect
them orthogonally must converge into the future. Horizons are then taken as the boundaries of regions of trapped surfaces
and are necessarily marginally trapped --- the expansion of ingoing null geodesics is negative but that of outgoing sets is zero.  
Such ideas are not new. The definition of trapped surfaces actually preceded that of event horizons \cite{penrose} 
and these objects are at least as characteristic of black holes  : they are clearly a measure of strong gravitational
fields, they imply the existence of spacetime singularities (event horizons do not), and they even imply the existence of 
event horizons in spacetimes with the appropriate causal structure.

What has changed over the last decade has been the increased interest in these ideas. This has partly been 
motivated by theoretical concerns that the identification of physical objects shouldn't depend on the structure
of null infinity and events which might or might not happen in the far future. However, at least as strong a
reason has come from numerical relativity. There, black holes are widely identified with outermost 
marginally trapped surfaces \cite{thomas}. Event horizons can be (approximately) located, but only 
retroactively after a simulation is complete. This de facto characterization, along with the theoretical 
concerns, was the original impetus  for studying these objects. As we shall see however, the rapid
progress in our understanding of quasilocally defined horizons now gives its own motivation to this work 
--- more and more they are seen as interesting and physically important objects in their own right. 

Studies of quasilocal horizons go by several names: in numerical work outermost marginally trapped surfaces
are usually referred to as apparent horizons (which does not correspond to the technical definition of that 
term), while the best known of the mathematical programmes are trapping \cite{hayward}, 
isolated \cite{isoreview}, and dynamical horizons \cite{ak}. In this paper we review the developments of the last
decade, seeking to clarify the nomenclature (often similar concepts have many 
names while sometimes one name is
used to refer to several distinct ideas), build an intuition about the physical relevance of the various objects, and understand
their many properties.

The paper is structured in the following way. To set the stage, in section \ref{horizons} we review the 
classical definitions of black holes including trapped surfaces and event, Killing, and apparent horizons.
Section \ref{qh} discusses some of the newer definitions such as isolated, trapping and dynamical horizons and their
relationship with those older ideas, in particular emphasizing their kinship to apparent horizons and trapped surfaces. 
Apart from definitions in principle however, we will also seek to test them by considering how they apply to various
sample spacetimes. This will help to illuminate their strengths and weaknesses and point the way to further investigations 
and improvements.  Then, 
in section \ref{props} we will explore the mathematical implications of these definitions, considering theorems on their
existence, uniqueness, allowed topologies, associated flux laws, limiting behaviours, and role as boundaries in 
Hamiltonian formulations of general relativity. 
In section \ref{sum} we will summarize, draw some conclusions, and look towards the future.

Finally, I give advance notice that this review will only really be of the classical uses 
and implications of quasilocal horizons. We will say very little about the quantum applications and 
so neglect such important developments as the role of isolated horizons in the loop quantum gravity 
entropy calculations \cite{entropy} and the developing applications of these ideas to quantum mechanical
investigations of black hole formation and evaporation \cite{evap}.

\section{Standard black holes and their boundaries}
\label{horizons}

\subsection{Event horizons}
\label{eh}

\begin{figure}[t]
\begin{center}
\input{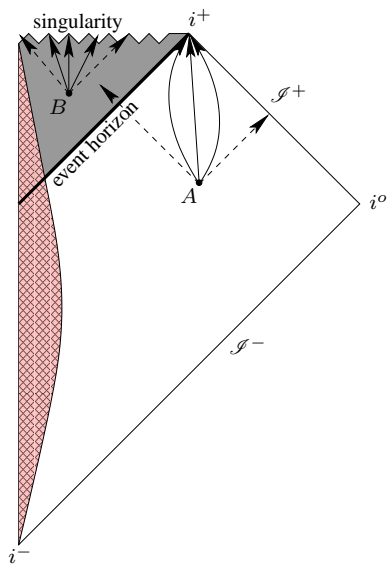}
\end{center}
\caption{A Penrose-Carter diagram showing a spacetime in which a matter distribution collapses to form
a black hole. As usual in these diagrams, null curves are drawn with slopes of $\pm 1$. 
$i^\pm$ is future/past timelike infinity (the destination/origin of timelike curves as 
``time" goes to $\pm \infty$), $i^o$ is the corresponding spacelike infinity, and $\mathscr{I}^{\pm}$
is future/past null infinity (the destination/origin origin of null curves as ``time" goes to $\pm  \infty$).
Note that point $A$ can send signals to $\mathscr{I}^+$ while $B$, which is part of the black hole, cannot.}
\label{EH}
\end{figure}

We begin with a quick review of event horizons, restricting our discussion to strongly asymptotically 
predictable spacetimes. The interested reader is directed to \cite{hawkellis, wald} for a proper 
definition of these terms (or \cite{chrusciel} for a recent discussion of associated mathematical 
difficulties)  but here we just note that they mean that the spacetime is asymptotically 
identical to Minkowski space. In particular, with the help of conformal mappings, one can rigorously
construct boundaries ``at infinity"  that correspond to the ultimate destinations/sources of (most)
inextendible spacelike, timelike, and null curves. As shown in Fig.~\ref{EH} these are
two-dimensional spacelike infinity $i^o$, the similarly two-dimensional past and future
timelike infinities $i^-$ and $i^+$, and the three-dimensional past and future null
infinities $\scri^-$ and $\scri^+$. Then a null curve is said to have ``escaped to infinity" if its counterpart
in the conformally mapped spacetime terminates on $\scri^+$.

In Minkowski space, all inextendible null curves terminate on $\mathscr{I}^+$. By contrast, spacetimes are said
to contain a (causally defined) \textit{black hole} if they have regions from which no null curve reaches $\scri^+$ --- thus
no causal curve can escape those regions. Equivalently, as shown in Fig.~\ref{EH}, 
a spacetime contains a black hole if the complement of the causal past of $\scri^+$ is non-empty. 
Then, that complement is the black hole and the boundary of that region is the \textit{event horizon}. 
In this example, null curves don't all escape to $\scri^+$ and those that don't end up either 
in the central singularity or trapped forever in the event horizon. 

\begin{figure}[t]
\begin{center}
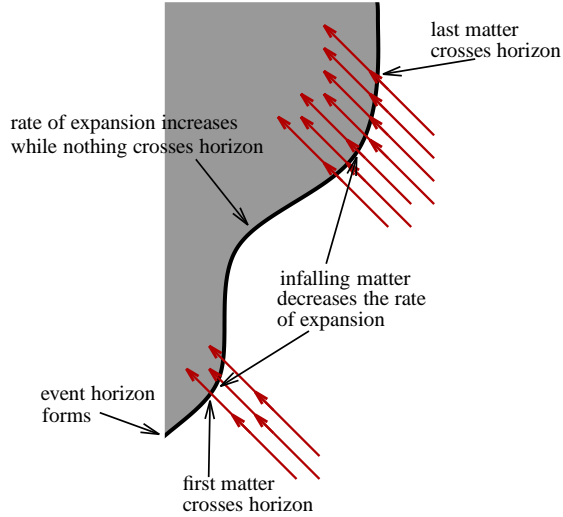
\end{center}
\caption{A two-dimensional schematic of the evolution of the event horizon (the heavy black line) as two concentric, spherically symmetric shells of matter (the sets of parallel arrows) collapse to form a black hole (the grey region). As noted in the text, the evolution is highly non-causal: the horizon forms in vacuum and then expands continuously with infalling matter \textit{decreasing} the rate of expansion. The expansion only stops when the last shell crosses the the horizon. In this diagram the horizon is non-expanding (and so null)
when it is vertical. Otherwise it is expanding.   }
\label{EHex}
\end{figure}

This definition is based on the global causal structure of the spacetime rather than directly on its 
local geometry/gravitational field. Further, it is teleological as the identification of a black hole
region depends on the ultimate fate of null curves, not their local behaviour. As mentioned in the introduction, these dual non-localities
mean that black holes evolve in non-intuitive ways. For example, one would naively expect that 
the event horizon of a black hole would expand as matter falls through it and be quiescent when 
it is (temporarily) isolated from such all influences. This is not
the case. As shown in Fig.~\ref{EHex}, infalling matter doesn't cause such expansion but instead
reduces rates of growth. By contrast, the absence of matter results in an acceleration of expansion. 
This growth only stops once all interactions \textit{permanently} cease between the event horizon and
its surroundings. Temporary cessations are not sufficient; due to the teleogical nature of the definition a 
black hole can distinguish between temporary and permanent states of isolation and react accordingly. 

To understand this behaviour keep in mind that the event horizon is null and so coincides with a 
congruence of null geodesics. As such its evolution is governed by the Raychaudhuri equation.
Thus (see for example \cite{eric})
\be
\frac{d \theta_{(\ell)}}{d \lambda} 
= -\frac{1}{2} \theta_{(\ell)}^2 - \sigma_{(\ell) ab} \sigma_{(\ell)}^{ab}
- 8 \pi T_{ab} \ell^a \ell^b \, , \label{dtl}
\ee
where $\ell^a$ is a null tangent vector to the congruence, $\lambda$ is an affine parameter for these geodesics, $\theta_{(\ell)}$ is the expansion of the geodesics (so that $d \vS / d \lambda = \theta_{(\ell)} \vS$, where $\vS$ is the area form), $\sigma_{(\ell) ab}$ is the shear, and $T_{ab}$ is the stress-energy tensor. 
The immediate thing to notice is that the first two terms on the right-hand side of the equation will always
be non-positive and, if the null energy condition holds, then so will the last term. Therefore, for any congruence
of null geodesics (including event horizons), $d \theta_{(\ell)}/d\lambda \leq 0$, and in particular it is not hard to 
show that if we ever have $\tl < 0$ then it must always diverge to $- \infty$ shortly thereafter. This result is the \emph{focussing theorem}; 
if a congruence of light
rays ever begins to contract in area then it must form a caustic  in finite time. Thus, for
an event horizon that asymptotes to become stationary we must always have $\tl \geq 0$. In turn this gives
rise to the celebrated area increase theorem  ($d \vS / d \lambda \geq 0$) for event horizons.

We can gain further understanding of this evolution by making the following, simple, calculation : 
\be
\frac{d^2 \vS }{d \lambda^2}  =  \left( \frac{1}{2}  \tl^2  - \sigma_{(\ell) ab} \sigma_{(\ell)}^{ab}
- 8 \pi T_{ab} \ell^a \ell^b \right) \vS \, . \label{ddA}
\ee
Thus, even though $\tl$ never increases, the rate of expansion of the area of the horizon can 
increase (unless $\tl = 0$), and does if $\tl \neq 0$ and the shear and matter terms vanish.  To clarify this difference,
consider the case of a sphere of light expanding outwards in Minkowski space. As it gets 
very large $\tl \rightarrow 0$ as local sections of the sphere become almost flat. By contrast, as this happens $dA / d \lambda \rightarrow \infty$ (where $A$ is the cross-sectional area of the horizon). 

With these observations in mind we can understand Fig.~\ref{EHex}. There, two spherical shells
collapse to form a black hole. Combining our omniscience with the symmetry, it is relatively easy to 
identify the event horizon. Specifically, outside the outermost shell the spacetime will be Schwarzschild
and static. Thus, once the two shells have collapsed within their (combined) Schwarzschild radius
we know that the event horizon will coincide with the surface $r=2M$. Then, we can trace the evolution
of this surface backwards in time to find its orgin at $r=0$. We note that there is nothing special happening
at $r=0$ at that point. Instead the event horizon appears in anticipation of future events (and at this 
point $\tl$ will have its maximum value). Having identified the relevant congruence 
the above equations (simplified by the symmetry so that $\sigma_{(\ell) ab} = 0$ ) 
guide its evolution as we follow it forwards in time. The newly formed horizon exists in vacuo and
so by (\ref{ddA}), the rate of growth of $A$ is positive and accelerating. This continues until the 
first shell of matter crosses the horizon and causes that rate to decrease until, in our example, it is
very close to zero. However, the horizon is then back in vacuum and so the growth of $A$ again
begins to accelerate. Just before the second shell hits, this rate is large enough to be noticeable and
we again see a rapid growth in the horizon area. Then, the second shell arrives, sends $\tl \rightarrow 0$, 
and so with no more matter ever falling into the hole, $A$ achieves its maximum value and stays
that way for the rest of time. 

Thus, there are at least three unusual features of event horizons (as compared to the 
usual objects studied by physics).  First, a standard black hole is only properly defined for spacetimes that
have a $\mathscr{I}^+$. This means that it is a highly non-local object as its very existence depends
on the structure of infinity. Second, even if the spacetime has the correct asymptotic structure we must
still wait until ``the end of time'' to locate the event horizon. It cannot be identified by local measurements. 
Third, while the evolution of the horizon is causal once it has been identified, its growth 
still reflect its initial, teleological identification. As such we have features like growth not being simply 
related to the rate of infalling matter. 

Of these, we have already seen that the first two are probably not causes for serious concern. 
Keeping in mind that this is an idealized
definition, it is perfectly reasonable to argue that in physical situations one should replace
$\scri^+$ with a set of observers, far from the black hole, who are willing to wait for long (but finite)
periods of time. Thus, as long as there is an ``almost" flat region around the black hole in which 
these observers can live, one can adapt this definition in an appropriate way (again see 
\cite{chrusciel}). That said, this clearly does not
resolve the problem for highly dynamical situations where such a region does not exist. Further, even 
if one can find a $\scri^+$ substitute, one still has to use non-local information to identify an event horizon
and live with its strange modes of evolution. 

Now, given the exotic nature of black holes it is certainly conceivable that this situation cannot be
improved upon; perhaps reality doesn't allow for a local identification of black holes. However, as we shall
see in the next couple of sections this does not seem to be the case. Alternative characterizations of black holes do exist and are, in general, at least as useful as causally defined black holes and event horizons. We now consider some of these
ideas. 


\subsection{Killing horizons}
\label{kh}
The best known examples of event horizons are in static or stationary spacetimes (ie. those in the Kerr-Newman solutions). 
As such, besides being null surfaces they are also tangent to one or more (global) Killing vector fields. This observation
gives rise to the following definition.
\\ \\
\textbf{Definition:} Given a spacetime $(M, g_{ab})$, a \emph{Killing horizon} is a null three-surface $H$ 
which is everywhere tangent to some Killing vector field $\xi^a$ of the full spacetime, which itself
is null on $H$ ($\left. \xi \cdot \xi \right|_H = 0$).  
\\ \\
It can be shown (see \cite{waldLR} for a general discussion or \cite{hawkellis} for details) that event
horizons in stationary, asymptotically flat, black hole solutions to the Einstein-Maxwell equations must be Killing 
horizons.  For static spacetimes (such as Schwarzschild), $\xi^a$ is the Killing vector field corresponding to 
time translations at infinity, but for stationary spacetimes (such as Kerr), it will be a linear combination of 
the ``time-translation" and rotational Killing vector fields : $\xi^a = (\partial/\partial t)^a + \Omega (\partial/\partial \phi)^a$,
where $\Omega$ is the angular velocity of the horizon. 

If one slightly relaxes the definition so that $\xi^a$ does not have to be a Killing vector field of the full spacetime, but instead
just of some neighbourhood of (some section) of $H$ then one can use Killing horizons to characterize horizons that
are (possibly temporarily) in equilibrium with their surroundings. Of course in this case they will not, in general, 
exactly coincide with the event horizon but in many cases they would be quite close (think about Fig.~\ref{EHex} where
the event horizon is ``almost" stationary). Then, on extracting a surface gravity $\kappa_{\xi}$ from the geodesic equation 
$\xi^a \nabla_a \xi^b =  \kappa_{\xi} \xi^b$, one can prove a version of the zeroth law of black hole mechanics : $\kappa_{\xi}$ is 
constant over $H$. The specific value of the surface gravity depends on the normalization of the Killing vector field, though 
if $\xi^a$ is a global Killing vector field, the natural choice is 
made by requiring that $\xi \cdot \xi = -1$ ``at infinity''
\cite{waldLR, eric}. 

Note that by definition the intrinsic and extrinsic geometry of $H$ will be invariant with respect to 
translations along the null Killing vector field. Thus, given a foliation of $H$ consistent with the Killing vector fields,
an induced metric $\tq_{ab}$ on the two-surfaces, and setting $\ell^a = \xi^a$ we have
\bea
\Lie_\ell \tq_{ab} = \tl \tq_{ab} + \sigma_{(\ell)ab} = 0 \, , 
\eea
and so the expansion and shear of the $\ell^a$ congruence vanishes. 

Finally note that while event horizons are necessarily Killing horizons under the conditions discussed above, the
converse is note necessarily true. For example, in Minkowski space with the standard rectangular coordinates,
the $x = \pm t$ three-surfaces are Killing horizons. Less trivially, 
\cite{lewanIso} constructs examples of other black hole-free spacetimes that can be entirely 
foliated by Killing horizons, each with topology $S^2 \times [0, 1]$. Thus, while Killing horizons may be characteristic
of stationary black holes they are not a sufficient condition for their identification. Situations of this type will recur with some of our
later definitions and the implication is simply that certain non-black hole structures share some common properties with black holes.

\subsection{Trapped surfaces and apparent horizons}
\label{trap}

For the Kerr-Newman family of black hole spacetimes, the black hole region has an alternative characterization that is
both local and physical.  Specifically, through each point inside the event horizon there exists a 
\textit{trapped surface} --- a closed two-surface which has the property that the expansions in each 
of the two forward-in-time pointing, normal-to-the-surface, null directions
$\ell^a$ and $n^a$ are everywhere negative.  That is, for a trapped two-surface $S$,
\begin{eqnarray} \tl = \tq^{ab} \nabla_{a} \ell_{b} < 0 \mspace{4mu} \mbox{ and }
\mspace{4mu} \tn = \tq^{ab} \nabla_{b} n_{b} < 0 \, ,  \label{TScond} 
\end{eqnarray}
where $\tq_{ab} = g_{ab} + \ell_a n_b + \ell_b n_a$  is the two-metric induced on $S$.
Here and in the future we always ``cross-normalize" so that 
$\ell \cdot n = -1$ and assign $\ell^a / n^a$ to be outward/inward pointing null vector fields
whenever it makes sense to make this distinction. 

One can compare such behaviour with, say, a two-sphere embedded in Minkowski space. 
There the ingoing set of null rays would focus ($\tn < 0$)  while the outgoing set would expand ($\tl > 0$). This is what one would intutively expect. By contrast, for trapped surfaces, the strong curvature of spacetime 
means that any causal evolution of the two-surface will be a contraction. 

Even away from these examples, there are general theorems that tell us that 
trapped surfaces characterize the strong gravitational fields intuitively associated with black holes. 
For example, for spacetimes which satisfy certain positive energy conditions and which have a $\scri^+$, any weakly trapped surface (where one only restricts both expansions to be non-positive) must be contained 
inside a black hole \cite{hawkellis, wald}.  Even for spacetimes without a  $\scri^+$, Israel \cite{israel} 
has shown that given the weak energy condition, any trapped two-surface 
can be extended into a spacelike three-surface of constant area that causally seals off its interior from the rest of the spacetime. 
Thus, trapped surfaces really are trapped. Finding a trapped surface in a spacetime is also sufficient to imply the
existence of a singularity  (in the form of an inextendible geodesic) somewhere in its 
causal future \cite{penrose, hawkellis, wald}. We note that there is no corresponding theorem for event horizons 
--- event horizons can exist that do not contain singularities. 

The association of trapped surfaces with black holes, forms the basis of the definition of 
an \textit{apparent horizon}.  
Following \cite{hawkellis}, if a  spacetime can be foliated by asymptotically flat spacelike three-surfaces 
$\Sigma_t$, then a point $q \in \Sigma_t$ is said to be \textit{trapped} if it lies on some trapped two-surface in $\Sigma_t$. 
The apparent horizon in $\Sigma_t$ is the boundary of the union of all of the trapped points and given certain smoothness assumptions,
it can be shown that it is a surface on which $\tl = 0$ \cite{hawkellis, wald, kriele}. Further, if the spacetime is strongly 
asymptotically predictable then the apparent horizon is contained inside (or coincides with) an event horizon. 

While it is certainly possible to identify an apparent horizon without reference to the far future,
this definition is quite unwieldy, most obviously in the need to check the assumption about
asymptotic flatness and also in the procedure of finding the boundary of the union of all the
trapped points. There can also be problems associated with the smoothness of the total trapped
region and/or the apparent horizon \cite{chrusciel}. In particular these problems mean that, contrary to 
popular belief (and usage), the outermost $\tl =0$ two-surface in a slice $\Sigma_t$ \textit{does not}
 always coincide with the original definition of an apparent horizon that we have just discussed. 
 
 That said, the outermost $\tl=0$ surface in a slice is obviously much more convenient to work with 
 than either an apparent or event horizon. In fact, as suggested above, the classical definition of
 an apparent horizon is widely ignored in practical applications such as numerical relativity
\cite{thomas} and instead the term is used to refer to the outermost $\tl=0$ surface.
This mixing of distinct ideas under the same name is nomenclaturally unfortunate but physically
suggestive. The quasilocal horizon definitions that we now consider can be thought of as
formalizations of this vernacular notion of an apparent horizon.

\section{Quasilocal horizons}
\label{qh}

For purposes of this paper event, Killing, and apparent horizons represent the classical notions of black hole
boundaries, and with this background in mind we now turn to the alternative, quasilocal definitions that
have been developed over the last decade. Based on the discussion of apparent horizons above, it won't be a 
surprise that each of these is defined in terms of $\tl = 0$ surfaces. However, we will also see that such surfaces
are not exclusively restricted to black hole spacetimes. Thus, further restrictions are needed to single out 
these surfaces as black hole boundaries. Again following the model of apparent horizons, we will see that requiring that 
there be trapped surfaces ``just inside" the horizon seems to be what is needed. 

For a quick reference, these definitions and their associated acronyms are summarized in Table
\ref{gloss}. 

\begin{table}
\begin{tabular}{|l|l|c|c|c|c|l|}
\hline
& &  & & & Signature  & \\
Acronym & Full name & Dim & $\tn$ & $\Lie_n \tl $  & (if 3D) & Other \\
\hline
IH & isolated horizon & 3 & - & - & null &  energy condition \\
MOTS & marginally outer trapped surface & 2 & - & - & - & closed topology \\ 
MOTT & marginally outer trapped tube & 3 & - & - & - & foliated by MOTSs \\ 
MTS & marginally trapped surface & 2 & $< 0$ & - & - & closed toplogy \\ 
MTT & marginally trapped tube & 3 & $< 0$ & - & - & foliated by MTSs \\
TH & trapping horizon & 3 & $\neq 0$ & $\neq 0$ & - & foliated by MOTSs \\
FITH & future inner trapping horizon & 3 & $<0$ & $>0$ & - & foliated by MTSs \\ 
FOTH & future outer trapping horizon & 3 & $<0$ & $<0$ & - & foliated by MTSs\\ 
DH & dynamical horizon & 3 & $<0$ & - & spacelike & foliated by MTSs \\
TLM & timelike membrane& 3 & $<0$ & - &  timelike & foliated by MTSs\\ 
\hline
\end{tabular}
\caption{A tabular summary of the various quasilocal horizons and their definitions. Note that other
properties may follow from those that are listed, but only requirements that explicitly appear in the definition have been included here. }
\label{gloss}
\end{table}

\subsection{Isolated horizons}
\label{ih}
We begin with isolated horizons. These are intended to be the black hole analogues of isolated,
equilibrium states in thermodynamics and so correspond to black holes that are not interacting with 
their surroundings, even though those surroudings themselves may be dynamic.
The simplest way to introduce these is as a generalization of Killing horizons.
While a Killing horizon assumes the existence of a Killing
vector field in some neighbourhood of $H$, an isolated horizon is defined only in terms of the  
intrinsic and extrinsic geometry of $H$ itself. As such, the existence of an isolated horizon does not 
assume any properties of the surrounding spacetime and in particular allows for dynamic evolutions of
the surrounding spacetime ``right outside" as long as the horizon itself is left unaffected. 

They have been well studied and are the subject of an extensive literature (see for example \cite{iso1, isoreview, isomechanics1, isomechanics2}). 
Here we will just review some of their main properties, and note that 
technically we'll be considering a variety known as weakly isolated horizons. 
\\ \\
\textbf{Definition:} A \textit{weakly isolated horizon} is a three-surface $H$ such that :
\begin{enumerate}
\item $H$ is null, 
\item the expansion $\tl = 0$ where $\ell^a$ is a null normal to $H$, 
\item $-T_a^b \ell^a$ is future directed and causal, and 
\item $\Lie_\ell \omega_a = 0$, where $\omega_a = - n_b \nabla_{\! \! \underleftarrow{a}} \ell^b$ and the
arrow indicates a pull-back to $H$.
\end{enumerate}
Note that the second condition does not depend on the specific normalization of $\ell^a$ since 
$\theta_{(f \ell)} = f \tl$ for any function $f$. The third is an energy condition (and in particular is implied by the
dominant energy condition). The last condition can always be met by an appropriate choice of $\ell^a$ and so
isn't a restriction on which surfaces may be considered to be weakly isolated but instead is just a rule 
about how we should choose the normalization of $\ell^a$. This restriction is non-unique. 

If we consider any foliation of $H$ into two-surfaces $S_v$ it is easy to see that the first two conditions
guarantee that $\Lie_\ell \vS = 0$ (where, as before, $\vS$ is the area form for the horizon) while from the Raychaudhuri equation (\ref{dtl}), we also have 
$\sigma_{(\ell) ab} = 0$ and $T_{ab} \ell^a \ell^b = 0$. Then, it follows that $\ell^a$ is a Killing
vector for the intrinsic geometry of $H$ and so this geometry is invariant under ``time evolutions" with
respect to $\ell^a$. Further, there are no matter flows across $H$. An isolated
horizon does not interact with its surroundings. 

That said those surroundings can, themselves, be dynamic. 
We have not imposed direct restrictions on the rest of the spacetime or even on a neighbourhood of $H$ as
we did for Killing horizons. No assumptions have been made about the behaviour of extensions of $\ell^a$ off of $H$
and in fact it has been shown \cite{lewanIso2} that the spacetime surrounding an isolated horizon does not have
to inherit any of the symmetries of $H$. In particular, there can exist spacetimes which are both dynamic and non-spherical
 ``right outside" a spherical isolated horizon. Certain Robinson-Trautman spacetimes are commonly cited as examples of these \cite{RT}. 
 
 Further support for isolated horizons as equilibrium states comes from a study of isolated horizon mechanics. 
 First, since $\ell^a$ generates a congruence of null geodesics that rule $H$, we know that
 there is a function $\kappa_\ell$ such that $\ell^a \nabla_a \ell^b = \kappa_\ell \ell^b$. As for Killling horizons,
 this is the \emph{surface gravity} of the horizon and we note that we can equivalently write 
 $\kappa_\ell = \ell^a \omega_a = - n_b \ell^a \nabla_a \ell^b$. Then, for any weakly isolated horizon it 
 can be shown that each choice of $\ell^a$ that satisfies the fourth condition of the definition gives rise to a 
 $\kappa_\ell$ that is constant everywhere on $H$ \cite{isomechanics1}.  This is the zeroth law of isolated horizon mechanics. Note that unless the isolated horizon is also a Killing horizon (and so we can normalize $\ell^a$ versus 
 times translations at infinity) significant freedom remains in the choice of $\ell^a$ and therefore the surface gravity. 
 In fact, while the formalism tells us that any $\kappa_\ell$ will be constant over $H$, it doesn't fix that constant. 
In many  circumstances there will be a convenient choice of that value, but that is extra input into the formalism. 

There is also a first law of isolated horizon mechanics. Here, we will restrict our
attention to (uncharged) weakly isolated horizons that also admit a rotational symmetry generated by some
vector field $\phi^a$ (technically these are \emph{rigidly rotating horizons}). 
For more general cases see \cite{isomechanics1, isomechanics2, isoDistort}.
The first law then comes from a careful Hamiltonian analysis of the phase space of spacetimes which have an isolated horizon as a boundary. In the usual way, the Hamiltonian functional includes both bulk and boundary terms and when
evaluated on any particular solution to the Einstein equations, the bulk terms vanish. Conventionally the boundary
terms are then identified as the energies associated with those boundaries and that is what is done here. In that case, 
it can be shown that the evolution of such spacetimes admits a Hamiltonian formulation if and only if for variations through the
phase space of isolated horizons
\bea
\delta H_{isol} = \frac{\kappa_\ell}{8 \pi} \delta A + \Omega \delta J_\phi \, , \label{firstlaw}
\eea
where $H_{isol}$ is the contribution from the isolated horizon boundary, $A$ is the area of the 
two-dimensional cross-sections $S_v$ of the horizon, and 
\bea
J_\phi = \int_{S_v} \vS \phi^a \omega_a \, , 
\eea
is the corresponding angular momentum associated with the rotational symmetry. $H_{isol}$, the surface
gravity $\kappa_\ell$, and the angular velocity $\Omega$ are functions of $A$ and $J_\phi$ alone. 


As the zeroth law fixes the surface gravity to be a constant but doesn't fix the actual value of that constant, the first law 
fixes relationships that the quantities must satisfy but doesn't actually specify expressions $H_{isol}$, $\kappa_\ell$, 
or $\Omega_\phi$. That said, if one calibrates the expressions for surface gravity and angular velocity so that they 
take their regular values for the Kerr section of the phase space, then (\ref{firstlaw}) can be integrated to show that
$H_{isol}$ must also take its usual value \cite{isomechanics1, isomechanics2}.

Finally, before moving on to non-isolated horizons, we note that just as for Killing horizons, 
the phase space of isolated horizons includes more than 
just outer black hole horizons. 
For example, all horizons in the fully extended Reissner-Nordstr\"{o}m-deSitter spacetime
qualify as isolated horizons --- the outer black hole horizon, the inner Cauchy horizon, the corresponding 
horizons associated with white holes, and even the cosmological horizons. Further
since all Killing horizons are also isolated horizons, 
the examples of non-black hole Killing horizons discussed in 
section \ref{kh} are also examples of isolated horizons. 
Thus, while the isolated horizons conditions are sufficient to 
capture many of the properties of equilibrium black holes they are not exclusively black holes. 
In the next section as we classify (potentially) non-isolated horizons, 
we will also consider the extra conditions that are necessary 
to distinguish between the various sub-classes of horizons. 

\subsection{Classifying horizons}
\label{nihdef}

We now consider quasilocal definitions of black hole 
boundaries which will include non-equilibrium states. These definitions will be general enough to allow for
isolated black hole horizons as equilibrium states and yet also include extra conditions to identify and exclude
the non-black hole isolated horizons considered above. 

Following the terminology of \cite{gregabhay},
the primary building block of these surfaces is the \textit{marginally outer trapped surface} (MOTS): 
a closed, spacelike, two-surface for which the outgoing null expansion $\tl = 0$ (from here on
we will always assume that we can distinguish between ``inward" and ``outward" pointing 
vector fields). A three-surface (or indeterminate signature), $H$, which can be entirely foliated with MOTSs, $S_v$, will be referred to as a \textit{marginally outer trapped tube} (MOTT).  We immediately see that 
a null MOTT (that also satisfies the energy condition) is an
example of an isolated horizon, though with a particular foliation. Note, however, that the restriction that the
MOTSs be closed spacelike surfaces means that certain Killing horizons, 
such as the $x = t$ surfaces in Minkowski space, are not MOTTs. 

To further restrict the allowed structures that we will call horizons, 
it is common to demand that $\tn < 0$ so that ingoing null rays 
focus. Such behaviour would be intuitively expected (and removes the Killing horizons of \cite{lewanIso} 
from consideration) and also has strong physical implications. If $\tn< 0$ then a MOTS is called a 
 \textit{marginally trapped surface (MTS)}  \footnote{Note that ``marginally trapped surface" is used to mean different things in different places in the literature. In one common usage (\cite{wald} for example) the surfaces that
we have called weakly trapped (with $\tl \leq 0$ and $\tn \leq 0$) are instead referred to as
marginally trapped. We do not use it in that way here and instead follow, for example, 
\cite{eric} or \cite{gregabhay} where a marginally trapped surface is as defined above in the main text.} 
and in turn this will imply the existence of
an event horizon in suitable spacetimes (an MTS is weakly trapped). 
Following \cite{ak, gregabhay} we also define a 
\textit{marginally trapped tube (MTT)} as an MOTT  with $\tn < 0$. 

At first thought these two conditions might seem to be sufficient to identify a surface as a black hole boundary. 
However this is not the case and there are non-black hole spacetimes that 
contain MTTs. For example, Senovilla \cite{senoDyn} has shown that certain vanishing scalar invariant (VSI) 
spacetimes include MTTs (and in fact can be foliated by them) but do not contain any trapped surfaces or other
signatures of black holes. 

Along with the definition of apparent horizons, this example
suggests a third condition that must be satisfied by any three-dimensional tube that aspires to be viewed as a black hole 
boundary --- we require that there be trapped surfaces ``just inside" the MTT. This can be made mathematically
precise by demanding that with an arbitrarily small ``inwards" deformation, each $S_v$ can be turned into a 
fully trapped surface. This idea was first proposed by Hayward \cite{hayward} and he 
formulated the condition in the following way (though using different notation). Construct a dual-null foliation of some neighbourhood
of the MTT so that, in particular, the MTSs are surfaces of constant null variables \cite{haywardDual}. Further, 
scale the null vectors $\ell^a$ and $n^a$ so that they generate the foliation (note that in general this will
mean that it is no longer possible to require $\ell \cdot n = -1$). Then, $\tl$ and $\tn$ are defined at every point
in the neighbourhood as the expansions of surfaces of constant null coordinates and if $\Lie_n \tl < 0$,
those surfaces ``just inside" the MTT are fully trapped. Conversely if $\Lie_n \tl > 0$ those ``just outside" 
are trapped (a situation that we will see in \ref{mtt} when we study a class of spacetimes that include
``jumping" MTTs) and if the sign of $\Lie_n \tl$ varies over an $S_v$, neither set is trapped. 

Of course such conditions have physical implications. As we saw in section \ref{trap} trapped surfaces
imply the existence of spacetime singularities. Thus any MTT that satisfies this condition almost certainly 
should be viewed as associated with a black hole --- it bounds trapped surfaces, conceals a spacetime singularity, 
and is contained in an event horizon (if the global structure is such that one may be defined). These arguments were
first made by Hayward and form the basis of his classification of MOTTs \cite{hayward} :
\\ \\
\textbf{Definition:} An MOTT is a \emph{trapping horizon} if $\tn \neq 0$ and there exists a dual-null 
foliation of some neighbourhood of $H$ so that $\Lie_n \tl \neq 0$. 
If $\tn < 0$ then a trapping horizon is \textit{future} while if $\tn > 0$ it is \textit{past}. Further, if 
$\Lie_n \tl < 0$ it is \textit{outer} while if $\Lie_n \tl > 0$ it is \textit{inner}. 
\\ \\
The names follow from a consideration of the various horizons of the fully extended Reissner-Nordstr\"{o}m solution. In that case, 
using the natural spherically symmetric dual-null foliation, it is easy to see that the black hole event and Cauchy horizons 
are future trapping horizons while the corresponding white hole horizons are past. 
Further, the black hole Cauchy horizon is a future inner trapping horizon (FITH) while the event horizon is a future outer trapping horizon (FOTH). 

Following \cite{ams05} the trapped surface condition may be reformulated in the following, computationally more convenient way. 
Instead of considering dual-null foliations, one directly considers the deformations of the MTSs generated by vector fields. Thus, given
a vector field $X^a$ defined over an $S_v$ one constructs an extension of $X^a$ over some neighbourhood of the
two-surface and then evolves that surface using the flow generated by the vector field (that is Lie-drags $S_v$ with $X^a$). Then one can
construct quantities such as $\tl$ on the new surface and so calculate its \textit{variation} : $\delta_X \tl$. As would be expected, 
the variation doesn't depend on the particular extension of $X^a$ and, for example, if $X^a = A \ell^a + B n^a$ we find
\bea
\delta_X \tl & = &   d^2 B - 2 \tom^a d_a B + (- d_a \tom^a + \tom_a \tom^a  - ^{^{(2)}} \! \! \! R/2 + 8 \pi T_{ab} \ell^a n^b) B \\
& & -A (\sigma_{(\ell) ab} \sigma_{(\ell)}^{ab}  + 8 \pi T_{ab} \ell^a \ell^b) \nn \, ,  \label{varTL}
\eea
where $d^2$ and $d_a$ are respectively the intrinsic Laplacian and induced covariant derivative on $S_v$ while
$\tom_a$ is the pull-back of $\omega_a$, and $^{^{(2)}} \! \! R/2$ is the two-dimensional Ricci scalar (this may most
easily be seen through an inspection of the relevant Newman-Penrose equations \cite{chandra}). Thus, the classification 
of a MOTT as a trapping horizon becomes a matter of classifying the ($A=0$) solutions to this differential equation
(again see \cite{ams05} for a related discussion). 

For spherically symmetric slices of Reissner-Nordstr\"{o}m horizons and the same symmetry for $B$, 
\bea
\delta_{(B n)} \tl = B \delta_n \tl =  B (- ^{^{(2)}} \! \! R/2 + 8 \pi T_{ab} \ell^a n^b) \, , 
\eea
and it is fairly easy to see that this matches the results of Hayward's classification. On the outer horizon, 
$\delta_n \tl = 0$ if and only if the black hole is extreme. Equally clearly though, if one breaks the spherical symmetry of $B$ we will not necessarily always have
$\delta_n \tl < 0$ --- a ``proper" scaling of the $n^a$ is crucial to the classification. A good example of the problems that this can cause arises 
if we consider black holes distorted by strong external gravitational fields \cite{distort, isoDistort}. There spherical symmetry is broken 
and if one does a naive calculation using surfaces of constant time coordinate to foliate the horizon, 
it does not appear that this is a FOTH. A more careful analysis is needed to properly classify 
these horizons.\footnote{As far as the author is aware, this has not yet been done.} 

Assuming that one can identify a FOTH, it is clearly a good candidate for an outer black hole boundary. That said, it would certainly 
be convenient to have a more direct way to identify these structures. It turns out that this is possible if we consider the 
\textit{dynamical horizons}  (DHs) of Ashtekar and Krishnan \cite{ak}. These are spacelike MTTs 
--- intuitively if isolated horizons are null one
would expect dynamic and expanding MTTs to be spacelike. As we shall see in section 
\ref{flux} this extra condition is sufficient to allow one to generate flux laws that track how the local flow of quantities across the dynamical 
horizon affects such properties as their total area. For now though we note that the VSI examples of \cite{senoDyn} again 
demonstrate that the spacelike
condition is not quite enough to identify a black hole boundary (those examples include dynamical horizons). However, if one also 
requires a genericity condition :  $\sigma_{(\ell) ab} \sigma_{(\ell)}^{ab}  + 8 \pi T_{ab} \ell^a \ell^b \neq 0$ plus the
null energy condition, then it can be shown that in this case FOTHs and DHs are equivalent  (section \ref{expansion} or \cite{gregabhay}).

To gain further insight into the physical relevance of these definitions we now turn to some examples of spacetimes which include
dynamically evolving horizons which may be confidently identified with black holes. 

\subsection{Dynamically evolving horizons}


As we have already noted, the quasilocal notions of horizons discussed above formalize the numerical relativity practice of 
identifying the outermost MOTS in a spacelike three-surface as a potential black hole boundary. 
Identifying this boundary is obviously physically interesting, but it also has practical advantages.
As we have seen, if the spacetime is asymptotically flat then any marginally trapped surface must 
lie within any event horizon, and so the interior of such a region may be safely excised without 
affecting the future development of the spacetime. 

Thus, if one consults almost any numerical study of black hole evolutions (eg.~\cite{numerExamples}), examples of dynamical
MOTT or MTT evolution may be found. The ``apparent horizon" is sometimes seen to evolve 
smoothly in a spacelike manner and sometimes seen to discontinuously ``jump" as new horizons
form around old. Unfortunately, given the involved nature of many of these simulations, 
complications arising from slicing conditions, and the fact that the interior of the ``apparent horizon" 
is generally discarded it can be difficult to use such simulations to build an intuition as to how 
these quasilocal horizons generically evolve. 

As such, exact analytic solutions, though necessarily much simpler than full numerical examples, play an important
role in building an intuition about how these boundaries evolve and which are relevant for black hole formation and
growth. In this subsection we'll consider in some detail how MTTs evolve in Tolman-Bondi spacetimes. It turns out that
while these are quite simple exact solutions, their MTTs demonstrate surprising rich behaviours. 

\begin{figure}[t]
\begin{center}
\input{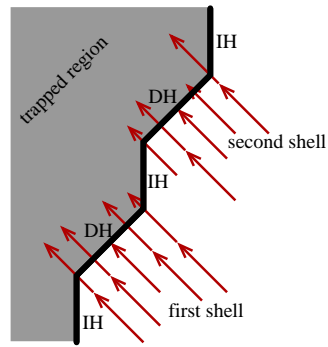}
\end{center}
\caption{A schematic of a Vaidya  spacetime in which two concentric shell of null dust collapse onto a pre-existing isolated horizon.
Isolated horizon sections are labelled IH while dynamic horizon (and so spacelike) regions are labelled DH. Note the difference
between the behaviour of the MTT versus the event horizon shown in Fig.~\ref{EHex}. }
\label{Vaidya}
\end{figure}

Before turning to these examples however, we briefly review two even simpler spacetimes and consider the questions that they
raise. To start, we consider what are probably the best known 
examples of exact dynamical black hole spacetimes --- the Vaidya solutions. These 
represent spherically symmetric black holes absorbing infalling null dust (thus the stress energy tensor for the 
spacetime takes the form $T_{ab} = \rho n_a n_b$ where $\rho$ is the dust density). Given the spherical 
symmetry one can easily identify MOTTs in these spacetimes and find that $\tn < 0$ and $\delta_n \tl < 0$. Further they are either null and isolated if not matter is falling through them 
or spacelike and expanding (dynamical horizons)
otherwise (Fig.~\ref{Vaidya}, \cite{eric, ak}). Thus, this example
supports the  intuition that dynamic MTT evolutions are generically spacelike. 

\begin{figure}[t]
\begin{center}
\input{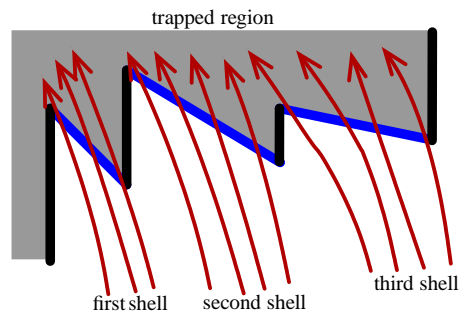}
\end{center}
\caption{An Oppenheimer-Snyder spacetime in which three consecutive shells of constant density dust fall into a pre-existing black hole. 
When no dust crosses the MTT, it is an isolated horizon while when it is dynamic it is a timelike membrane. Note that the trapped region may 
be either inside or outside of the MTT depending on whether it is dynamic or isolated.}
\label{OS}
\end{figure}

Howevever, this initial expectation is confounded by our second example --- 
Oppenheimer-Snyder collapse \cite{bendov}. These spacetimes are generated using spacetime surgery
to piece together sections of Schwarzschild and Robertson-Walker spacetimes to model the gravitational collapse of 
balls and/or shells of uniform density dust. MTTs exist in these spacetimes  and are null and isolated
if no matter is falling through them, but timelike and shrinking otherwise; these surfaces, often referred to 
as \emph{timelike membranes} (TLMs), connect isolated horizons either to each other or to the central density singularity
as shown in Fig.~\ref{OS}.  Equivalently $\delta_n \tl < 0$ in isolation (so that trapped surfaces are inside the horizon) but in 
dynamical situations $\delta_n \tl > 0$ (so that the trapped surfaces lie outside the TLM). 

Now, while the first example shows a smooth expansion of the horizon 
it appears that the second may be an example
of the ``jumping" apparent horizons seen in numerical relatlivity. The question then arises as to which 
of these behaviours 
is generic? In general do black holes expand smoothly or do the horizons jump? 
As neither of these examples is completely
satisfactory physically (null dust is very unusual matter while the second example has density 
discontinuities as well as other
unusual features), it is hard to draw general conclusions about which represents ``most" black hole evolutions. 

To better understand these evolutions we now turn to examples that are physically slightly more 
satisfactory --- spherically symmetric, but radially non-homogeneous, timelike dust collapse.

\subsubsection{MOTS in Tolman-Bondi spacetimes}
\label{mtt}

From the Tolman-Bondi spacetimes we can draw a whole series of physically interesting, but still fairly simple examples of 
spherical dust collapse. These were recently investigated in some detail in \cite{mttpaper} and here we review some of the
results from that paper. As already mentioned these solutions  
model the gravitational collapse of spherically symmetric but radially non-homogeneous clouds of dust. 
They are commonly used to study shell-crossings and the consequent naked singularities (see for example, the discussion in \cite{gonc}).  
However, as we shall now see, they also provide many interesting examples of MTT spacetimes.

The combination of pressureless dust and spherical symmetry makes the evolution of these spacetimes 
especially easy to study using only analytical tools. The spherical symmetry means that the problem
reduces to a consideration of the evolution of dust shells while the nature of this matter means that those
shells interact with each other only through the gravitational field. In turn, this means that tracing the evolution of each dust shell is essentially equivalent to calculating the evolution of radial geodesics in Schwarzschild spacetime. 

We study these spacetimes by giving initial conditions on a spherically symmetric, spacelike, 
three-surface $\Sigma_o$. We then specify a dust density $\rho_o$ so that on that
surface $T_{ab} = \rho_o \hu_a \hu_b$, where $\hu_a$ is the forward-in-time pointing unit normal to 
$\Sigma_o$. For purposes of this review we will also assume that the spacetime (and this matter)
is instantaneously static at $\Sigma_o$; that is, the extrinsic curvature of $\Sigma_o$ vanishes. 
Then the three-metric on $\Sigma_o$ is 
\bea
ds^2 = \frac{dr^2}{1 - 2 m(r)/r} + r^2 (d \theta^2 + \sin^2 \theta d \phi^2) \,  ,
\eea
where $r$ is the areal radius (so that a sphere with coordinate radius $r$ on $\Sigma_o$ 
has area $4 \pi r^2$), $\theta$ and $\phi$ are the usual spherical coordinates and 
\be
m(r) = \int_o^r d\tilde{r} (4 \pi \tilde{r}^2) \rho_o(\tilde{r})\, , 
\ee
is a mass function --- if the mass distribution were to terminate at $r$ then $m(r)$ would be equal to the
ADM mass of the entire spacetime. 

From these initial conditions the Einstein equations can be solved to evolve the full spacetime. Then in
Gaussian normal coordinates where $\tau$ is the proper time as measured by a fleet of geodesic
observers with initial velocity $\hu^a$ on $\Sigma_o$ (and so comoving with the dust):
\be
ds^2 = - d\tau^2 + \frac{(R'(\tau,r))^2}{1 - 2m(r)/r}dr^2 
+ R^2(\tau,r)\, d \Omega^2 \, ,
\ee
where $R(\tau,r)$ is the areal radius at time $\tau$ of the spherical set of observers who had initial areal radius
$r$, and must be a solution of 
\be
\dot{R}(\tau,r) \equiv \frac{dR(\tau,r)}{d\tau} = 
- \sqrt{ \frac{2m(r)}{R(\tau,r)}}   \sqrt{1 - \frac{R(\tau,r)}{r} }\, . 
\label{EErem}
\ee
$R(\tau,r)' = \partial R(\tau,r) / \partial r$ and the stress-energy tensor keeps 
the form $T_{ab} = \rho(\tau,r) \nabla_a \tau \nabla_b \tau$.
All of these quantities may be obtained analytically since (\ref{EErem}) may be parametrically
integrated. Details may be found in \cite{mttpaper}.
%

Since our spacetime is spherically symmetric, we search for MOTS which share this symmetry. 
Focussing our attention of these surfaces one can show that, up to multiplication by positive functions
of $r$ and $\tau$ : 
%
\be
\tl = \frac{2 ( \dot{R}(\tau,r) + \sqrt{1-2m(r)/r} )}{R(\tau,r)}  
\, \, \, \mbox{and} \, \, \, 
\tn = \frac{\dot{R}(\tau,r) - \sqrt{1- 2m(r)/r} }{R(\tau,r)} \, .
\ee
Then, consulting (\ref{EErem}) we then find that $\tl = 0$ if and only if $R(\tau,r) = 2 m(r)$, and on that surface we always have
$\tn < 0$. 
Thus, all quasilocal horizons in this spacetime are marginally trapped tubes (or future trapping horizons whenever
$\Lie_n \tl \neq 0$). 

\subsubsection{Examples}

\begin{figure}[t]
\begin{minipage}{6cm}
\resizebox{6cm}{6cm}{\includegraphics{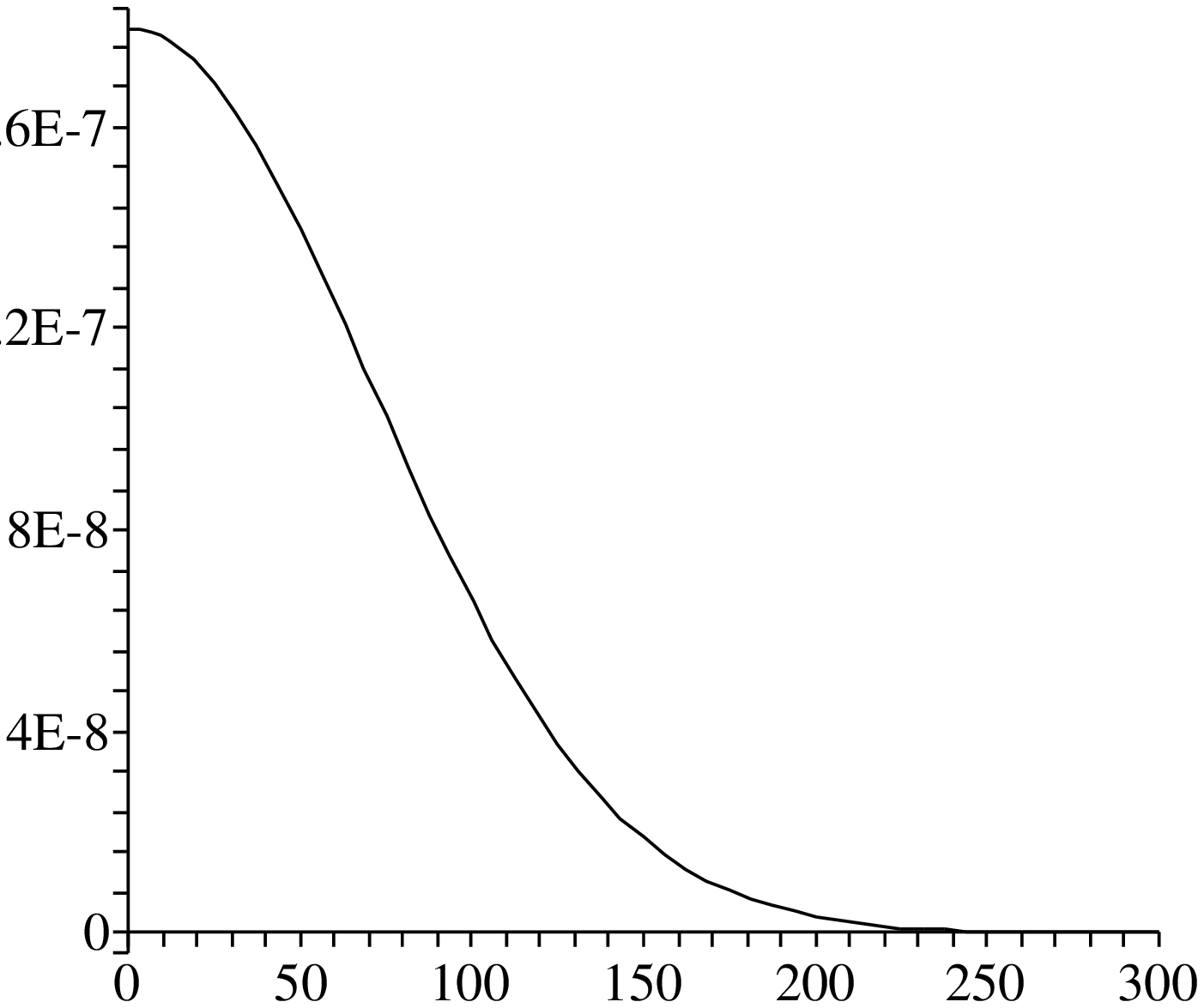}}
\put(-110,-5){a) $\rho_o$ vs $r$}
\end{minipage}
\begin{minipage}{6cm}
\resizebox{6cm}{6cm}{\includegraphics{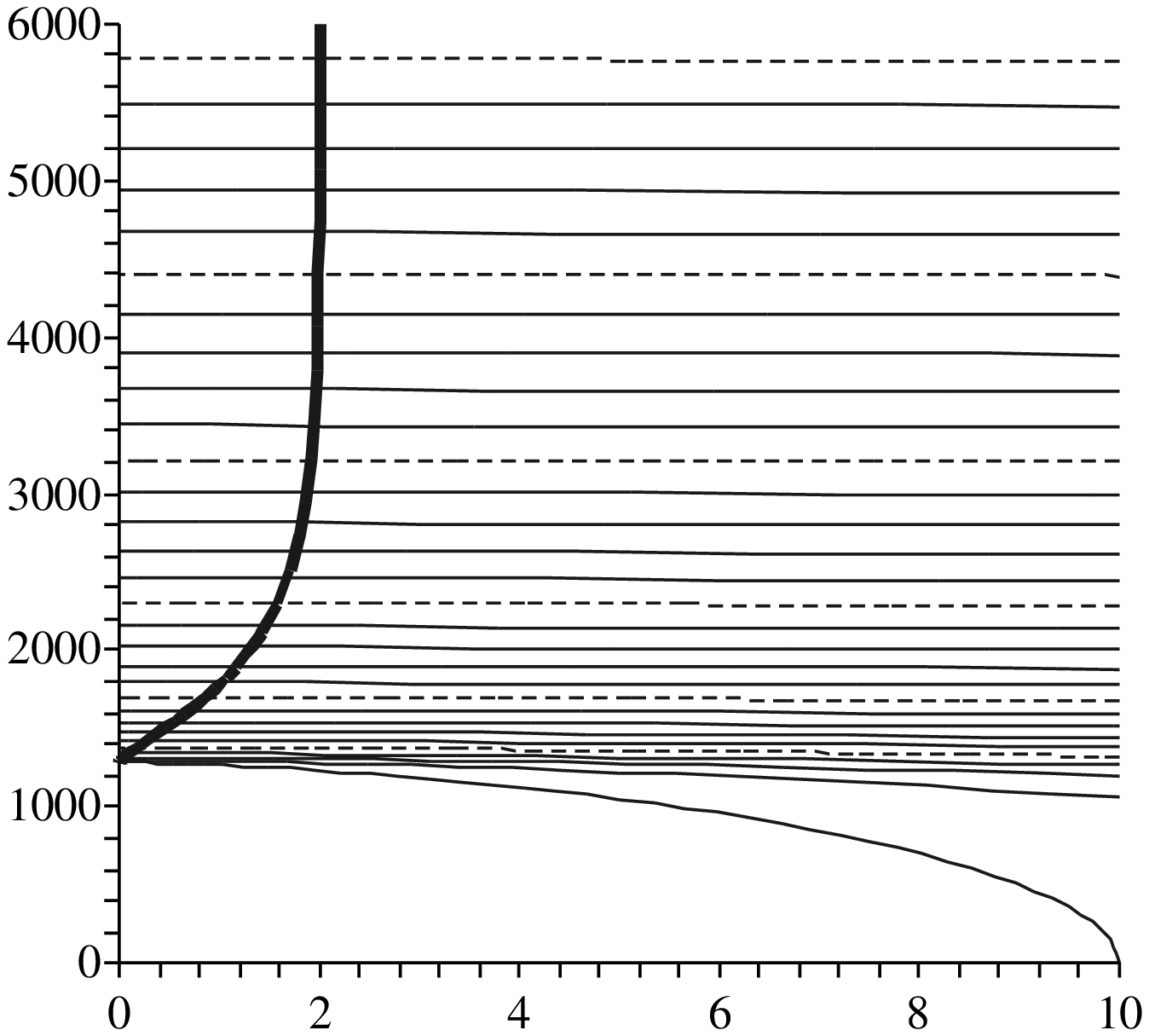}}
\put(-140,-5){b) $\tau$ vs $R$ for MTT evolution}
\put(-18,45){ {\footnotesize{$r=50$}}}
\put(-18,55){ {\footnotesize{$r=100$}}}
\put(-18,68){ {\footnotesize{$r=150$}}}
\put(-18,90){ {\footnotesize{$r=200$}}}
\put(-18,117){ {\footnotesize{$r=250$}}}
\put(-18,147){ {\footnotesize{$r=300$}}}
\end{minipage}
\caption{Collapse of a dust ball to form a black hole}
\label{dustcollapse}
\end{figure}

\begin{figure}[t]
\begin{minipage}{6cm}
\resizebox{6cm}{6cm}{\includegraphics{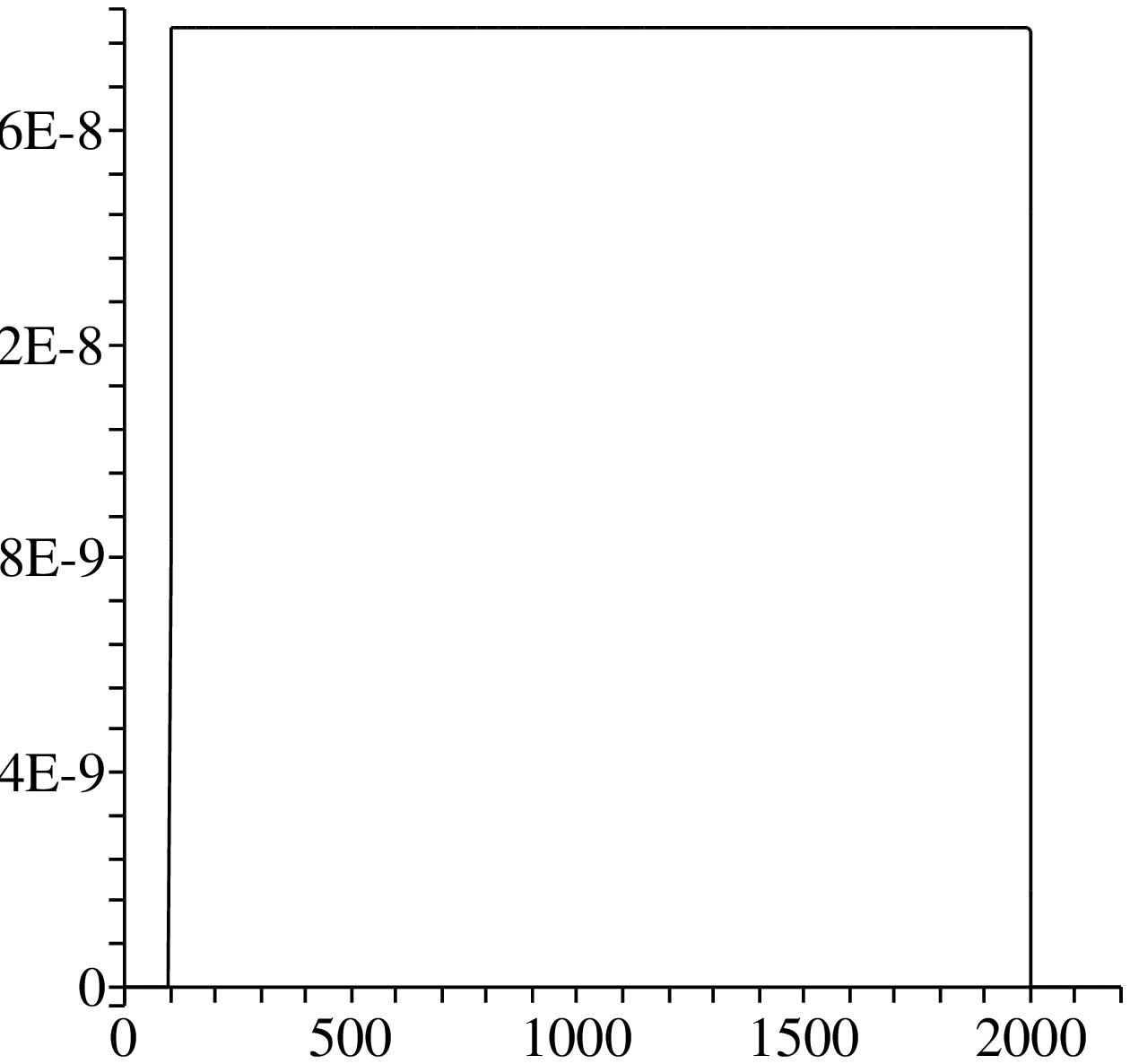}}
\put(-110,-5){a) $\rho_o$ vs $r$}
\end{minipage}
\begin{minipage}{6cm}
\resizebox{6cm}{6cm}{\includegraphics{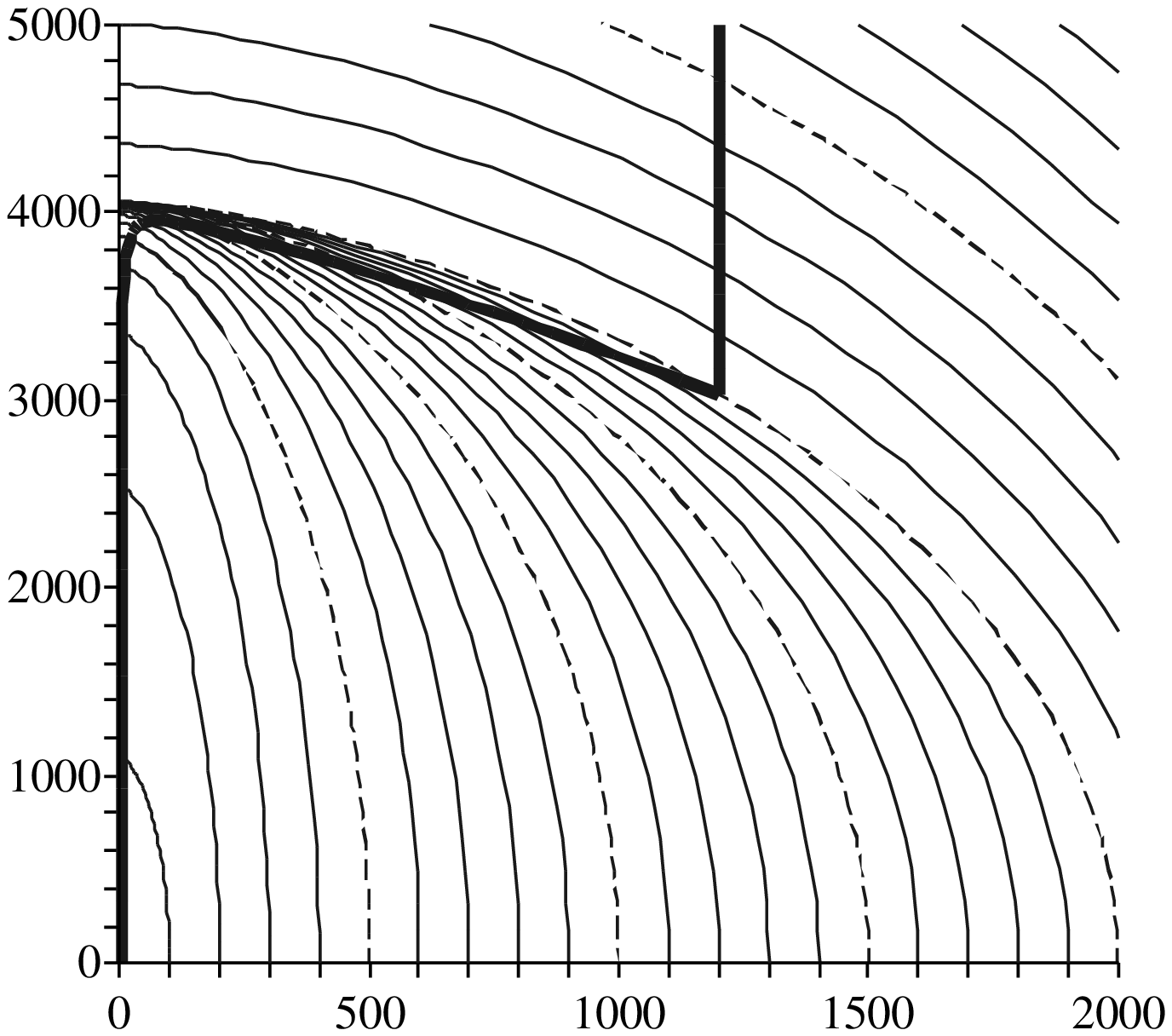}}
\put(-140,-5){b) $\tau$ vs $R$ for MTT evolution}
\end{minipage}
\caption{Large shell of constant density falls ``into" a small black hole}
\label{large}
\end{figure}

\begin{figure}[t]
\begin{minipage}{6cm}
\resizebox{6cm}{6cm}{\includegraphics{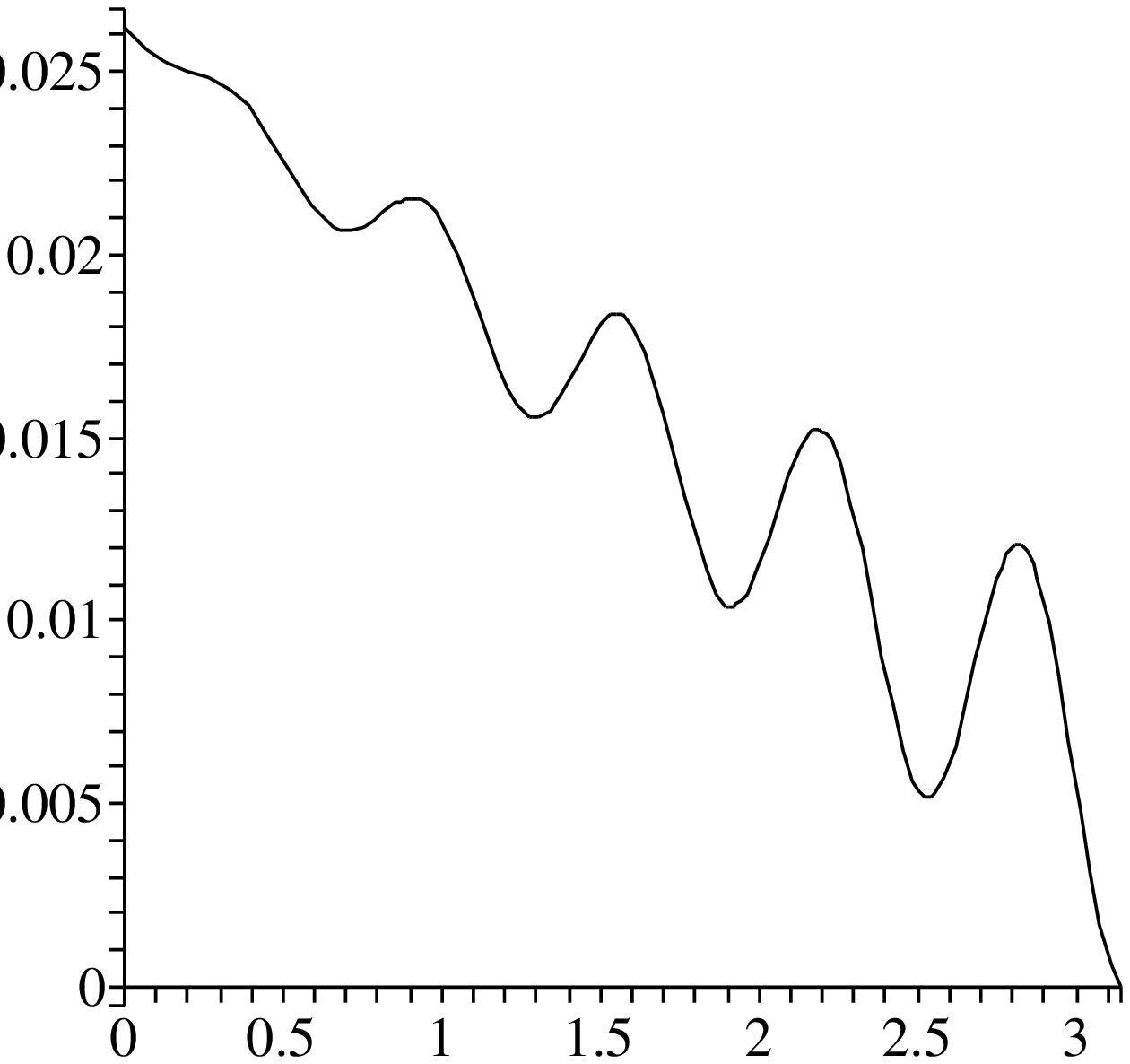}}
\put(-110,-5){a) $\rho_o$ vs $r$}
\end{minipage}
\begin{minipage}{6cm}
\resizebox{6cm}{6cm}{\includegraphics{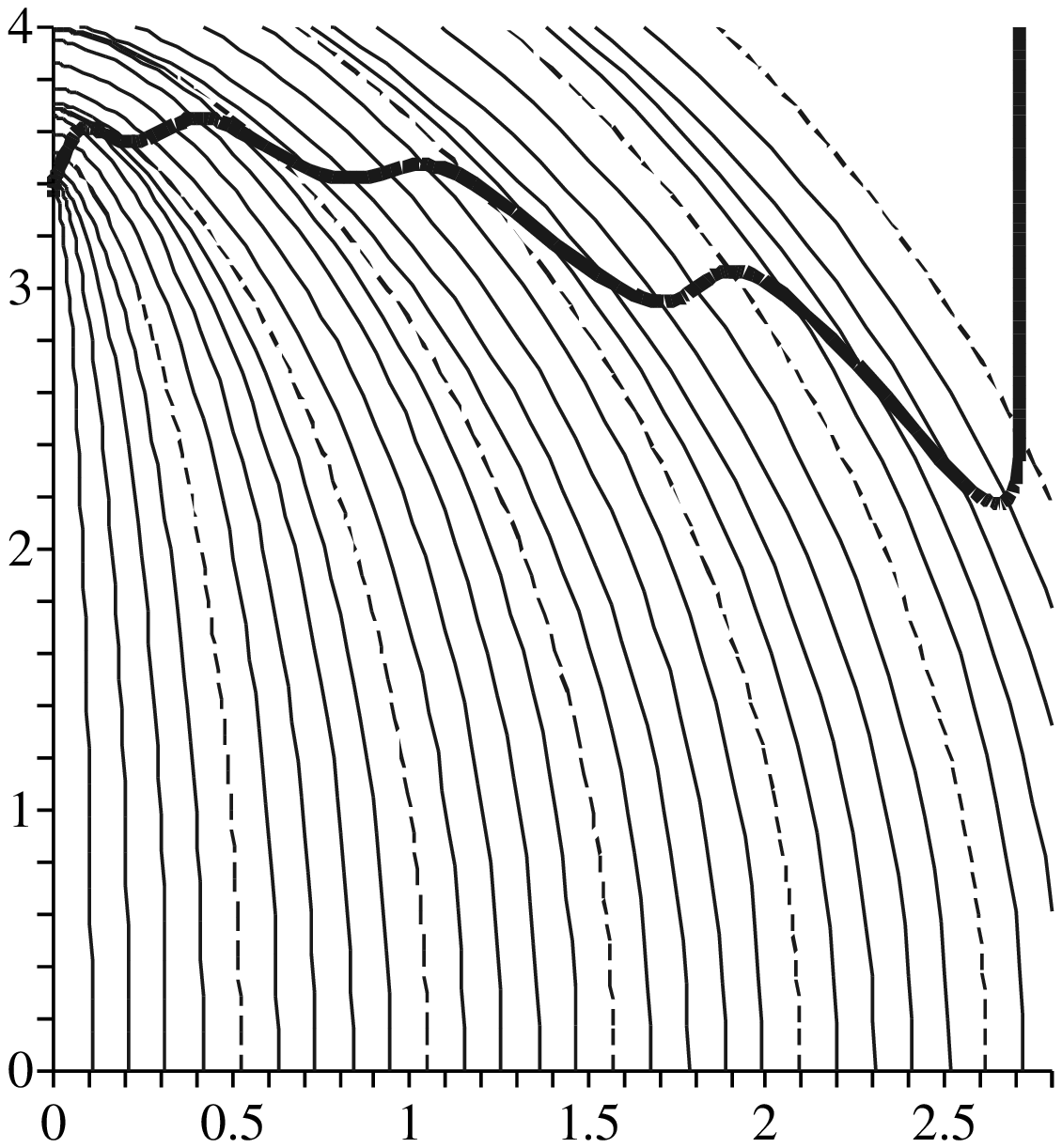}}
\put(-140,-5){b) $\tau$ vs $R$ for MTT evolution}
\end{minipage}
\caption{Complicated dust distributions collapse to form black hole}
\label{R3}
\end{figure}

Generating examples of MTT evolutions is then simply a matter of picking initial matter distribution functions $\rho_o(r)$ and evolving the data in time. Some care needs to be taken in picking initial
matter distributions to avoid shell-crossings but a great amount of 
freedom remains. Some examples are shown in Figs.~\ref{dustcollapse} - \ref{R3}.

Fig.~\ref{dustcollapse} shows the collapse of a spherical symmetric cloud of dust with an (initial) 
Gaussian density function. In that figure, all measurements are made relative to the ADM mass, $m_o$, of the
dust cloud. Part a) shows the initial mass distribution while part b) shows how the horizon (the thick 
black line) evolves as a function of $\tau$. The other lines in b) trace the paths of infalling dust shells.
Note that the horizon forms as the first shells collapse to zero area (and infinite density). It then 
grows from the singularity before asymptoting to $R = 2 m_o$ as most of the mass has fallen into 
the newly formed hole. Though we do not do so here, it is straightforward to show that the horizon 
is spacelike throughout this expansion and that $\Lie_n \tl < 0$. Thus, this is both a dynamical and
a future outer trapping horizon.

Fig.~\ref{large} shows another dramatic situation. In this case the initial configuration is a small 
black hole of mass $m_o$ surrounded by a huge cloud of constant density dust  with a total 
mass of $600m_o$.\footnote{Though initial black holes were not discussed in the above, they may 
easily be inserted into the initial data using spacetime surgery to excise and connect regions of
suitable spacetimes. See \cite{mttpaper} for the details. } Note that the density distribution is defined in 
terms of error functions and so, despite initial appearances, is smooth everywhere. 
The evolution of this distribution is shown in b) where we see
a slow initial growth followed by a ``jump" at $\tau \approx 3000m$ as a new horizon forms at 
$R \approx 1202$ around the old. It is fairly easy to show that the segment connecting the new 
horizon to the old is a timelike membrane. 

Alternatively, we can trace the evolution as a function of $R$. Then, it begins as an isolated horizon
(before the first dust shells hit), becomes a dynamical horizon (during the initial expansion), transitions
through a null (but not isolated) instant to become a timelike membrane (heading backwards in time), 
and then goes back to being dynamical before asymptoting towards isolation. Thus, 
it is a future trapping horizon everywhere except when it passes through
the null instants to transit from being a FOTH to being a FITH. During those instants $\Lie_n \tl = 0$. 

Finally, Fig.~\ref{R3} shows an example of a more complicated initial dust distribution and 
subsequent collapse. In this observers defining instants of simultaneity as surfaces of constant
$\tau$ would see new horizons form outside of old ones on four distinct occasions. In each 
case a DH/TLM (or FOTH/FITH) pair would appear with the DH (FOTH) growing and the TLM (FITH) 
shrinking. Further, on four separate occasions these shrinking ``horizons" would meet up with 
growing ones and disappear. Each of the creation/destruction events occurs on a surface of 
$\Lie_n \tl = 0$. 

Note that even though the MTTs are seen to become timelike in two of these examples and so may be crossed
by causally evolving observers, this does not mean that these TLMs may be used as escape hatches from the 
black holes. As is discussed in \cite{mttpaper} such observers may temporarily escape the trapped region but are always doomed
to be recaptured. 

It is also of some interest to consider the physical circumstances under which the horizon ``jumps" can occur. Again it is shown
in \cite{mttpaper} that these jumps occur when the infalling matter is dense relative to $1/A$, where $A$ is the surface
area of the horizon. For a solar mass black hole this means that jumps occur only when the infalling matter is at least as 
dense as a neutron star while it is only when one gets to galactic mass black holes that jumps could be generated by matter
with the density of water. This suggests an answer to the question posed in relation to the Vaidya and Oppenheimer-Snyder
spacetimes. Both evolutions can occur but ``jumps" only happen under very extreme conditions (for small black holes). Essentially
they correspond to new horizons forming outside of old ones. 

One of the surprising observations from these examples is that FOTHs/DHs do not necessarily last forever once they are 
formed. They may appear and disappear, though apparently according to strict rules which always 
leave the trapped region shrouded from outside observers. 
The formations/destructions are signalled by the presence of TLMs that connect new horizons to old. How far this
behaviour extends into non-spherically symmetric cases is currently unknown but in particular it seems possible that 
the multiple $\tl = 0$ surfaces sometimes seen inside a numerical ``apparent" horizon \cite{numerExamples} may actually represent one (or possibly a few) MTTs weaving backwards and forwards in time. We will come back to this point in section
\ref{sum}.

\section{Properties of quasilocal horizons}
\label{props}

We now review some properties that follow from the definitions set forth in the last section. 

\subsection{Existence and Uniqueness}

While it is clear from both numerical simulations and our examples that the quasilocal horizons discussed in the
previous section exist under a wide range of circumstances, there is surprisingly little known about what
mathematical conditions are needed to guarantee this existence. Slightly more
is known about the lack of uniqueness of apparent horizons/MOTTs. In particular, it is known that in general a MOTS can be
distorted and so any MOTT will be highly non-unique \cite{eardley}. Even more dramatically it has been shown 
that apparent horizons are completely absent from some slicings of Schwarzschild
spacetime \cite{wald1}. That said, recent work has shed new light on both of these questions and we now review these
results, starting with existence.

\subsubsection{Existence}
\label{exist}

Assume that a spacetime $(M, g_{ab})$ is foliated with spacelike hypersurfaces $\Sigma_t$ (with all fields and
surfaces smooth). Further assume that on a slice $\Sigma_o$ of that foliation, there exists a MOTS
and for some function $\beta$, the variation
$\delta_{ \beta \hat{r} } \tl > 0$ everywhere on $S_o$, where $\hat{r}^a$ is the outward pointing unit
normal vector to $S_o$ in $\Sigma_o$. Then, from our discussion in section \ref{nihdef}, there are trapped
surfaces in $\Sigma_o$ ``just inside" of $S_o$, and $S_o$ is said to be strictly stably outermost (SSO). 

It has very recently been shown \cite{ams05} that these conditions imply that 
$S_o$ is part of a marginally outer-trapped tube $H$ 
such that each leaf $S_v$ of $H$ lies in some leaf $\Sigma_t$ of the spacetime foliation. That is
$S_v = H \cap \Sigma_{t(v)}$ for some function $t(v)$.
Further, $H$ exists at least as long as the $S_v$ remains SSO and, if the
null energy condition holds, is spacelike if $\sigma_{(\ell) ab} \sigma_{(\ell)}^{ab}  + 8 \pi T_{ab} \ell^a \ell^b \neq 0$. 
Else it is null (and so isolated). Note that no condition has been imposed on $\tn$. 



Given this result, MOTSs will always evolve into MOTTs in numerical relativity at least until the SSO
condition is violated. Based on the Tolman-Bondi examples of the last section we already have a good idea of when 
such violations might occur. Examples of such violations were seen in Fig.~\ref{large} and \ref{R3} where the marginally trapped tubes
transition from being spacelike to being timelike and heading backwards in time (and exactly where $\delta_n \tl = 0$). In those situations
the MOTT will necessarily become (momentarily) tangent to any foliation of spacetime and so the existence theorem fails --- which is
a good thing since we have seen in these examples that the tube does not continue to smoothly evolve forwards in time relative to 
the foliation. Thus, the theorem is consistent with our examples.  

We note however that violations can also occur as a result of (unfortunate) foliation choices. Though 
we haven't discussed such examples here, reference \cite{mttpaper} shows that it is fairly easy to generate foliations that are 
(at least temporarily) tangent to a dynamical horizon and even interweave it. 
This is a function of a foliation choice rather than physical or mathematical necessity, but 
we note that in these cases the SSO condition (but not $\delta_n \tl<0$) would also be violated. That said,
in these examples, the MTT again fails to continue evolving 
``forward in time" relative to the foliation and so the existence theorem is still consistent.


\subsubsection{Uniqueness}
\label{unique}

There are two distinct types of uniqueness that may be considered with respect to quasilocally defined horizons. First, 
given a MOTT, is it possible to foliate it into MOTSs in more than one way? Second, given a spacetime with two or 
more MOTTs (produced, for example, by alternate foliations of the full spacetime) are there any restrictions on how 
they may be positioned relative to each other? A recent paper by Ashtekar and Galloway \cite{gregabhay} addresses both of these 
issues. 

The first issue is completely resolved for (globally) achronal FOTHs  (or equivalently achronal dynamical horizons that
satisfy $\sigma_{(\ell) ab} \sigma_{(\ell)}^{ab} + 8 \pi T_{ab} \ell^a \ell^b \neq 0$). In this case, it is shown that
the foliation of a FOTH is unique and so it is impossible to refoliate it in any other way. Thus, in these cases the foliation is
an intrinsic property of the hypersurface. This can be contrasted with the situation for isolated horizons.  No mention
of a foliation is made in the definition of those structures, neither is it needed since $\ell_a$ is defined as the
normal to $H$ as a whole rather than some two-surface. Indeed it is not hard to see that isolated horizons will
admit many possible foliations (and consequent definitions of $n_a$, the in-horizon normal to those two-surfaces). 
Note too that timelike membranes are not achronal either and so possible refoliations of them are not forbidden by the theorem. 

One immediate implication of this theorem arises if we consider the dependence of apparent horizons on spacetime foliations.
To see this, assume that we have a foliation $\Sigma_t$ of a spacetime $M$ and have used it to identify a compatible 
FOTH, so that $S_t = \Sigma_t \cap H$. Now let $\Sigma'_t$ be a second foliation so that the $S'_t = \Sigma'_t \cap H$ 
are not the same set of two-surfaces as the $S_t$ (ie. the foliations are not tangent at $H$). 
Then, by the uniqueness result, the $S'_t$ cannot be MOTS. Thus, any $H'$ identified with respect 
to the new foliation cannot coincide with $H$ and so, as expected, FOTHs determined in this way must vary
with the foliation of the spacetime. 

\begin{figure}
\input{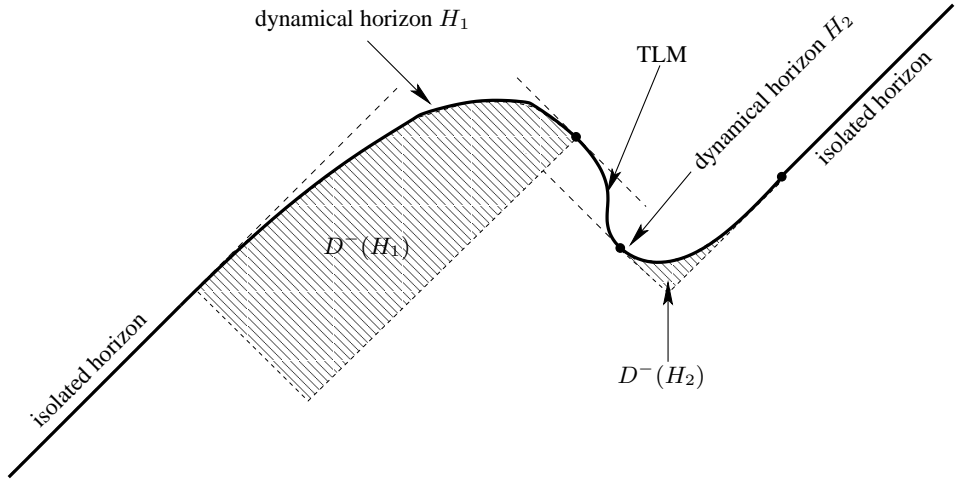}
\caption{A schematic showing an MTS which transitions from isolated-dynamical-TLM-dynamical-isolated. The large
dots indicate transitions points between these behaviours. Dashed lines are null surfaces and shading indicates
regions where the uniqueness theorems forbid the existence of WTSs.}
\label{AKfig}
\end{figure}

This paper also sheds light on the second issue. It is shown that, given the null energy condition and a regular
dynamical horizon/achronal FOTH $H$, no weakly trapped surface can be contained in $D^-(H)$, the past 
domain of dependence of $H$. An application of this theorem to a specific MTS with two regular dynamical horizon
regions is depicted in Fig.~\ref{AKfig}. There, the past domains of dependence are shown as shaded regions. 

There are several immediate implications of this result. First, it is clear that a spacetime cannot be foliated by regular
dynamical horizons. If this were possible then there would certainly be weakly trapped surfaces in the past of each horizon. 
In this context we note that the VSI examples considered earlier fail the 
$\sigma_{(\ell) ab} \sigma_{(\ell)}^{ab} +  8 \pi T_{ab} \ell^a \ell^b \neq 0$ condition and so are consistent with this
theorem. Note however that it does not forbid weakly trapped surfaces having components that lie in $D^-(H)$. 
For example it would be possible to have a second dynamical horizon $H'$ that ``interweaves" with the first 
so that each MOTS lies partly in the past and partly in the future of $H$. Thus, while multiple dynamical horizons 
are certainly allowed by this theorem, their behaviour is strongly constrained. In the example of horizons generated
by spacetime foliations, we then know that while different foliations will necessarily generate different horizons, those
horizons will just as necessarily be constrained to repeatedly intersect each other. 

Finally, before moving on, we will briefly discuss \textit{trapping boundaries} \cite{hayward}. It has been proposed that an invariantly 
defined trapping horizon could be defined in analogous way to how apparent horizons are obtained. 
Specifically, one would find the 
union of all of the points in a spacetime that lie on any trapped surface and so define a 
\emph{trapped region}. The three-dimensional boundary of this region is the trapping boundary and it was shown that, given
certain assumptions about the relationship between the trapped surfaces and some foliation of the trapping boundary,
it must be a trapping horizon. 
If these conditions are generically met, then 
this construction would give us a canonical trapping horizon. Unfortunately (see the discussion
of  wild MTSs in \cite{eardley} and the discussion in \cite{gregabhay}) it appears that that the assumptions
are probably not satisfied in general. 
Instead it seems likely (though it has not been proven) that the trapping boundary will actually turn out
to be the event horizon --- in general the wild MTSs will sample the far future and so can access the necessary 
information to locate the event horizon.

\subsection{Laws of expansion}
\label{expansion}

As was discussed in section \ref{eh}, one of the best known properties of classical, causally defined, black holes is that, given the null energy condition, they never decrease in area. It was first shown in \cite{hayward} this is also true for certain classes of quasilocal horizons. Using dual null foliations of the 
type discussed in section \ref{nihdef}, the argument is fairly straightforward though we will modify 
its statement a little here. 

Let $H$ be a FOTH and let 
$\cV^a \in TH$ be a tangent vector field that satisfies $\mathcal{V} \cdot n < 0$ and 
is both foliation generating and orthogonal to those foliation two-surfaces.
Then, if the null energy condition holds the \emph{area increase theorem} says that 
either : 1) $H$ is null and isolated or 2) $H$ is spacelike and 
$\Lie_{\cV} \vS \geq 0$. That is, under the flow generated by $\mathcal{V}^a$ the area element $\vS$ 
never decreases in magnitude and hence the same holds true for the area of the $S_v$ two-surfaces.  

In considering this result note that the $\cV \cdot n$ condition picks the orientation of $\mathcal{V}^a$ and in particular this ensures that it points 
forwards-in-time when $H$ is isolated. This, combined with the second two conditions means that for a given 
foliation parameter $v$, we may always choose the normalization of $\ell^a$ such that
$\mathcal{V}^a = \ell^a - C n^a$ for some function $C$ (which we'll refer to as the  
\emph{expansion parameter}) and $\Lie_{\mathcal{V}} v = 1$. 
The result then follows quite simply. Since $\tl =0$ everywhere on $H$ we must have
\be
\Lie_{\cV} \tl = \Lie_\ell \tl - C \Lie_n \tl = 0 \Leftrightarrow C = \frac{\Lie_\ell \tl}{\Lie_n \tl} \, .  \label{FOTHeq}
\ee
The division is always possible since $\Lie_n \tl < 0$ by assumption. Meanwhile by the Raychaudhuri 
equation (\ref{dtl}) and the null energy condition $\Lie_\ell \tl \leq 0$. Then $C \geq 0$ everywhere on $H$ and
$C=0$ if and only if the horizon is temporarily (or permanently) isolated. Further, note that
$\cV \cdot \cV = 2 C$ and so $H$ is either null (and thus an isolated horizon) or spacelike (and so a dynamical horizon). 
Finally, $\Lie_{\cV} \vS = - C \tn \vS$ where $\tn < 0$ by assumption and so the horizon never decreases in area. 
It is also clear that if the null-energy condition is violated (as it would be by Hawking radiation) then we could have
$\Lie_\ell \tl > 0$ and so a FOTH that decreases in area. 

From these considerations, it is clear that a dynamical horizon that satisfies the genericity condition 
(which may be essentially be rephrased as $\Lie_\ell \tl \neq 0$) is necessarily a FOTH; it is spacelike which means
that $C > 0$ and so, in turn, we must have $\Lie_n \tl < 0$. However, even without the genericity condition, $C>0$ is 
sufficient to ensure that a dynamical horizon expands in area (as indeed the non-black hole DHs do in the VSI 
examples \cite{senoDyn}). By contrast the strictly stably outermost MOTTs considered in section \ref{exist} do not
necessarily increase in area as there no assumption was made about the sign of $\tn$. 

Timelike membranes/FITHs necessarily decrease in area. At first this might appear to be 
a conundrum given some of the Tolman-Bondi examples which always seem to expand. 
The apparent difficulty is resolved when one realizes that the area increase theorem
 (reasonably) assumes that $\cV^a$ on a TLM will be oriented to
point forwards-in-time. With this orientation the TLMs seen in the examples do shrink and this is certainly a valid
way to view their evolution. If, however, one demands that $\cV^a$ be consistently oriented with a foliation parameter $v$
on the MTT, with $v$ always increasing forwards-in-time on isolated sections, then one naturally has $\cV^a$ pointing
backwards-in-time on TLM sections. Thus, in all of our examples we have $dA/dv \geq 0$. With this perspective
we can see that if an MTT interpolates between two asymptotically isolated regions, the later isolated horizon (relative to the foliation 
parameter) will always have a larger area then the earlier. 

\subsection{Topology}
In the preceding discussions we have often implicitly assumed a spherical topology for MOTS. 
In fact, for an outer trapping horizon which satisfies
the dominant energy condition, this is not an assumption but follows directly from the Einstein equations. 
To see this integrate equation (\ref{varTL}) over the MOTS with $X^a = n^a$ to obtain
\bea
 \chi = \frac{1}{\pi} \int_{S_v} \vS \left\{ -  \delta_n \tl + \tom_a \tom^a + 8 \pi T_{ab} \ell^a n^b \right\} \, ,  \label{top}
\eea
where we have used the well-known result that the integral of the Ricci scalar over a two-surface is the 
topological invariant $2 \pi \chi$,  
with $\chi$ equal to the Euler characteristic of the surface. 
Now, by the $\delta_n \tl < 0$ assumption and the dominant energy condition the right-hand
side of this equation is positive and so $\chi$ must be postive as well. However, the only closed and 
orientable two-surfaces with positive Euler characteristic are spheres. 
Thus, the two-surfaces foliating an outer trapping horizon (including FOTHs) must be spheres. 
This result was first shown specifically for trapping horizons in \cite{hayward} but has been known for some time for strictly stably outermost
MOTSs \cite{ams05, newman}.

It has been noted \cite{ak} that this result is left unchanged in the presence of a positive cosmological constant but fails for $\Lambda < 0$. In that case a negative term enters the right-hand side of equation (\ref{top})
and so it is no longer strictly positive. Then, alternative topologies are allowed
which is in accord with the fact that event horizons with such topologies can exist in asymptotically
anti-deSitter spacetimes \cite{Aminneborg:1996iz}.

\subsection{Flux laws}
\label{flux}

Given the area increase laws, in an ideal world we would also like to have a corresponding set of flux laws that
describe how incoming energy and matter flows generate those increases. A priori,  it is not at all
obvious that such laws could actually exist given that the process will, in general, be highly dynamical and so one might 
expect that it could be impossible to localize these contributions. Surprisingly, this does not seem to be the case. It has
been shown that for dynamical horizons \cite{ak} a whole range of such flux laws exist. This was extended to 
certain classes of FOTHs (and so transitions to and from isolation) in \cite{haywardflux} . Finally, results from 
\cite{ams05} appear to imply that this class may include all strictly stably outermost FOTHs. Thus, for a wide
class of horizons we have quasilocal flux laws that govern their growth.

Our review of these developments begins 
with the original result \cite{ak}. Given a dynamical horizon $H$, we can define its forward-in-time
pointing unit normal $\hat{\tau}_a$ to $H$ and a spacelike unit
tangent vector field $\hat{r}^a$ that is normal to each of the $S_v$ and whose orientation is chosen
by requiring that $\hat{r} \cdot n < 0$. Then, the flux laws originate from the following observation:
\bea
& & \mathcal{H} + 2 \hr^a \mathcal{H}_a   \label{akEq1} \\
& = & ^{^{(2)}}\mspace{-4mu} R - \sigma_{(\hat{\ell})}^{ab} \sigma_{(\hat{\ell}) ab}  
- 2 \zeta^a \zeta_a + 2 d_a \zeta^a 
 + \mbox{(terms that vanish when $\tl = 0$)} \, ,  \nn
\eea 
where $\mathcal{H}$ and $\mathcal{H}_a$ are the Hamiltonian and diffeomorphism constraints on $H$ and so 
satisfy $\mathcal{H} + 2 \hr^a \mathcal{H}_a = 16 \pi  \hat{\ell}^a T_{ab} \htau^b$,
while $\hat{\ell}_a = \htau_a + \hr_a$, and $\zeta_a =  \hat{r}^b \nabla_b \hat{\ell}_{\underleftarrow{a}}$ (where the arrow subscript
indicates a pull-back to $S_v$). 

Then, as for the topology calculation, we integrate (\ref{akEq1}) over $S_v$ (this time using the fact that $\chi = 2$)
to find that :  
\bea
8 \pi  = \int_{S_v} \vS \left\{  16 \pi T_{ab} (\htau^a + \hr^a) \htau^b 
+ \sigma_{(\hat{\ell})}^{ab} \sigma_{(\hat{\ell}) ab} + 2 \zeta^a \zeta_a \right\} \, . \label{id}
\eea
For any function $ 4 F(v)$ we can then use the fundamental theorem of calculus to write 
\bea
F(v_2) - F(v_1) =  \int_{H_{21}} \vH \left\{T_{ab} \xi_r^a \htau^b  
+ \frac{N_r}{16 \pi} \left(\sigma_{(\hat{\ell})}^{ab} \sigma_{(\hat{\ell}) ab} + 2 \zeta^a \zeta_a  \right) \right\} \, , \label{spaceflux}
\eea
where the integration is over some the section of a dynamical horizon bounded by $S_1$ and $S_2$, 
$\vH  = \pm \mathbf{\hat{r}} \wedge \vS$ is the natural volume form on this spacelike surface, 
$N_F = || \bdv || \frac{dF}{dv}$ and $\xi_r^a = N_r \hat{\ell}^a$. $N_F$ may be identified as a natural lapse function
associated with the function $F$ while $\xi_r^a$ is the corresponding (null) time vector field. 

Equation (\ref{spaceflux}) is the Ashtekar-Krishan flux law for any quantity $F(v)$ that may be defined on each $S_v$ 
expressing its change in terms of a matter flux across the $H$, 
a shear term (usually interpreted as a flux of gravitational wave-type energy) and the 
$\zeta^a \zeta_a$ term (whose meaning isn't so clear though see \cite{haywardWave} for one interpretation). 
Particularly interesting quantities include the irreducible mass $\sqrt{A/16 \pi}$ and  $A/4$ (in which case the flux law might be interpreted as measuring the flow of entropy).

In this original form there are difficultiies in taking the isolated limit --- many of the quantities are defined in terms of 
$\hat{r}^a$ and $\htau^a$ which in turn are only well defined for spacelike $H$. However, recent developments suggest that
this limit may be fully achievable. In \cite{haywardflux} it was pointed out that if one switches to the coordinate dependent 
volume form $\bdv \wedge \vS$ and defines
\bea
\tilde{\ell}^a = \sqrt{ \frac{dF}{dv} } \left( \htau^a + \hr^a \right) \; \; \mbox{and} \; \; 
\tilde{n}^a = \sqrt{ \frac{dF}{dv} } \left( \htau^a - \hr^a \right) \, , 
\eea
then we can rewrite equation (\ref{spaceflux}) as 
\bea
F(v_2)-F(v_1) 
= \int_{v_1}^{v_2} dv  \int_{S_v} \vS \left\{  T_{ab} \tilde{\ell}^a \tilde{\tau}^b
+ \frac{1}{16 \pi} \left( \sigma_{(\tilde{\ell})}^{ab} \sigma_{(\tilde{\ell}) ab} + 2 \tilde{\zeta}^a \tilde{\zeta}_a \right)  \label{fluxhay}
 \right\}  \, ,  \label{flux2}
\eea
where the various quantities have been rewritten in terms of $\tilde{\ell}^a$ and $\tilde{n}^a$. The various quantities will then 
remain well-defined in the isolated limit if $dF/dv \rightarrow 0$ sufficiently quickly to keep the new null vectors both non-zero and
non-divergent. This limiting process will be considered in more detail in \cite{SlowEv} (see also the discussion in \cite{ak}) but for now note 
that it only really has a chance of being well defined if any transition between dynamical and isolated regions happens ``all at once". That
is, for no $S_v$ is $H$ partly dynamical and partly isolated. Equivalently on a given $S_v$ either $C=0$ everywhere or nowhere. 
Now, while it might seem unlikely that such transitions would be generic, in fact \cite{ams05} proves just this result for
strictly stably outermost MOTTs. 

There are also flux laws for angular momentum \cite{ak}. We do not discuss them here but
note that they are essentially the same as those governing the angular momentum 
evolution of timelike tubes \cite{by};  they are not restricted to dynamical horizons and 
actually hold for any three-surface regardless of its signature. A nice discussion of these laws 
may be found in \cite{gourg}. There it is shown that the fluxes may be physically interpreted 
by viewing the evolving horizon as a viscous fluid and the flux laws as a version of the 
Navier-Stokes equations. 

\subsection{Slowly evolving horizons}

The limiting process considered above occurs in the regime of ``almost isolated" horizons. This is also the regime 
where one would expect the classical results of black hole perturbation theory to apply and would include many, if not most, 
astrophysical processes --- the exceptions being especially dramatic events like the final stages of black hole mergers. Finally,
one would expect that the black hole analogues of thermodynamic quasi-equilibrium states would be ``almost isolated", and so
physical process versions of the laws of black hole mechanics should apply here. Thus, an investigation of such states 
is of considerable physical interest.  A couple of approaches have been used to study this limit. Here we will focus on the that 
taken in \cite{prl, bill, SlowEv} though the interested reader is also referred to \cite{haywardPert} for an alternative. 

A \emph{slowly evolving horizon} is an ``almost isolated'' FOTH. Now, it is not immediately obvious that such a characterization
is possible. In particular, 
what does it mean for a spacelike (or timelike) surface to be ``almost" null and how can this be expressed
in an invariant manner? In \cite{prl} we proposed that if one regularizes $\ell^a$ against the FOTH foliation parameter $v$ 
so that
$\Lie_\cV v = 1$ (with $\cV = \ell^a - C n ^a$ as before), the key condition for identifying a FOTH as slowly evolving over some
slice $S_v$ is that 
\bea
\sqrt{|C|} |\tn| <   \frac{\varepsilon}{R} \, , \label{SEcond}
\eea
where $\varepsilon < < 1$ and $R$ is the areal radius of $S_v$. The first thing to note is that this expression is invariant under 
rescalings of $\ell^a \rightarrow \alpha \ell^a$, since in this case $C \rightarrow \alpha^2 C$ and $n^a \rightarrow \frac{1}{\alpha} n^a$.
Thus, this is a geometrically meaningful statement to make. Next it is clear that this parameter vanishes for isolated 
horizons. Finally, on a 
dynamical horizon where $\hr^a = \cV^a / |\cV|$ we note that  $\Lie_{\hr} \vS = \sqrt{C/2} \tn \vS < (\varepsilon/R) \vS$ and so in the approach 
to isolation our parameter is a measure of the geometrically invariant rate of change (and it goes to zero).\footnote{
There are also a few other technical conditions that a horizon must meet in order to be considered slowly evolving. 
These are chiefly restrictions on the environment surrounding the FOTH to ensure that it is not too extreme 
--- if this is the case then the horizon will not remain slowly evolving for very long. One suspects that they may also 
eliminate many, if not all, of the ``wild" MTSs discussed in section \ref{unique}.  } 

That said, the chief justification for viewing a FOTH which satisfies these conditions as slowly evolving comes from
its utility in calculations and its applicability to situations that one would intuitively consider to be quasi-equilibrium states.
Thus, it has been shown \cite{bill} that classical examples of perturbed black holes, such as a Schwarzschild black hole
immersed in a slowly changing tidal field \cite{tidal}, 
satisfy all of the requirements necessary to be classed as slowly evolving. A second example comes from considering 
the collapse of dust balls, such as those shown in Fig.~\ref{dustcollapse}. In those cases, it has been shown that while MTTs 
may not be slowly evolving just after formation, they rapidly asymptote to this condition as 
the amount of matter crossing the horizon also asymptotes to zero (again \cite{bill}). 

Thus, at least as much as is possible for a spacelike surface, 
it seems reasonable to view slowly evolving horizons as ``almost isolated".
Then, an analysis of the implications of these assumptions on the Einstein equations gives rise to other results that further
supports our interpretation. In particular, it can be shown that all other physical and geometrical quantities on the FOTH are 
also slowly evolving (this time relative to Lie derivatives with respect to $\cV$ which has been normalized so that 
$\cV \cdot \cV = 2 C \sim \varepsilon^2$). For example the surface shears slowly and any flow of matter across it is small. 
Further it can be shown that $\kappa_{\cV} = - n_b \cV^a \nabla_a \ell^b$ is constant (up to order $\varepsilon$) over an
particular $S_v$ and $\Lie_{\cV} \kappa_{\cV}$ is similarly small. Then, as this expression clearly reduces to the usual surface
gravity in isolation, it is natural to identify it as the approximate surface gravity here as well. 
Next, a careful analysis of the diffeomorphism constraint on $H$
shows that it reduces to:
 \bea
 \left(\frac{\kappa_o }{8 \pi}\right) \frac{dA}{dv} \approx  
 \int_{S_v} \vS \left( T_{ab} \ell^a \ell^b + \frac{1}{8 \pi } \sigma_{(\ell) ab} \sigma_{(\ell)}^{ab} \right) \, , \label{firstlawEH}
 \eea
 to lowest order in our approximation. We do not discuss how angular momentum terms may be included in this expression here. However
 they certainly can be and the interested reader is referred to \cite{prl, SlowEv}.
 
The analogy with quasi-equilibrium states in thermodynamics is clear. 
Just as such objects may be assigned a temperature, in 
this case we may assign a surface gravity to the hole. It is not completely constant and can slowly change 
but this is expected
behaviour for a system that is nearly in equilbrium with itself and its surroundings. 
Further, in the quasi-equilibrium regime of thermodynamics
one expects the physical process version of the first law to be expressible in the form $\dot{E} = T \dot{S} = \dot{W}$. 
This is exactly the form
seen in (\ref{firstlawEH}) if one identifies surface gravity with temperature and entropy with area in the usual way. 
Note too that the RHS also 
matches the form for energy flows across a perturbed event horizon first found in \cite{hawkFirst, tidal}. 

The precise relationship between (\ref{firstlawEH}) and the Ashtekar-Krishnan flux law (\ref{fluxhay}) 
is not immediately clear. For example, the null vectors appearing in the laws have very different behaviours in the 
isolated limit and there is no $\zeta^2$ term in (\ref{firstlawEH}). 
This will be discussed in detail in \cite{SlowEv}.

\subsection{A Hamiltonian formulation}

Both the flux laws and slowly evolving horizons are concerned with the exchange of energy and angular momentum between black holes
and their environment. That said in neither case is a specific expression for the energy singled out by the respective formalisms --- the 
flux laws can essentially be applied to any increasing function while the first law for slowly evolving
horizons  also allows for a rescaling freedom in 
$\ell^a$ (and therefore $\kappa_{\cV}$) essentially equivalent to the one allowed for isolated horizons. In an ideal world this freedom
would be reduced or eliminated and that is a major reason for examining these horizons from a Hamiltonian perspective. 

Such an analysis was conducted in \cite{thebeast}. The calculations are 
long and involved but the basic ideas behind them are
straightforward. In that paper,
 quasilocal horizons are examined from both Lagrangian and Hamiltonian perspectives. The 
Lagrangian approach is of the style used in \cite{by} and so begins by identifying 
a quasilocal action whose first variation vanishes 
for solutions to the Einstein equations that include one of these horizons as a boundary. 
A Hamiltonian functional is then derived using a Legendre transform and Hamilton-Jacobi arguments. 
By this process it is found that the action formulation is well-defined
if, on the horizon boundary, one sets $\tl =0$ 
as well as fixing the angular-momentum one-form $\tom_a$ and induced two-metric
$\tq_{ab}$ on the foliation two-surfaces (up to diffeomorphisms). 
Further, one has to fix the pull-back $\ell_{\underleftarrow{a}}$ to $H$ --- essentially
a gauge-fix so that specifying $\tom_a$ (the connection on the $S_v$ normal bundle) is physically meaningful. 
Thus, only a MOTT is needed to get a well-defined action principle and conditions on $\tn$ and the proximity of 
trapped surfaces are not necessary. 

From the analysis it turns out that only functionals of the fixed data $\tq_{ab}$, $\tom_a$, and $\ell_{\underleftarrow{a}}$  
are allowed as energies associated with the MOTT.  While this, unfortunately, does not completely eliminate the freedom found 
in the earlier analyses 
it does reduce it greatly while still allowing for most common energy expressions.
These conclusions are confirmed and re-examined using an extended phase space analysis in the style of \cite{bf1}. 
One advantage of these methods is they let one say something about angular momentum in addition to energy. 
In particular, in this case they uniquely select an expression for angular momentum and this expression 
precisely matches those appearing in the previous formalisms. 

Finally, largely based on the Hamiltonian perspective, it was seen that certain physical arguments allow one to say 
more about  energy and the flow of time. Specifically one requires that the expression for energy: 1)  be dimensionally 
correct, 2) be constant for isolated horizons, and 3) scale linearly with some notion of time. 
Then, based on the allowed freedom in the energy
expression one is lead to identify $\ell^a$ as the natural time vector field (which, in contrast to usual Hamiltonian formulations, 
does not generate evolutions of the boundary). A scaling of $\ell^a$ is forced by the
formalism and it turns out that this makes it essentially identical to the $\tilde{\ell}^a$ appearing in (\ref{fluxhay}). Finally, again 
based on the requirements, the allowed energy expressions are closely tied to $\ell^a$ and, though still not completely specified, 
shown to be constant in isolation and increasing on dynamical horizons (in agreement with the flux laws). It is argued that 
the irreducible mass $\sqrt{A/16\pi}$ is then a natural candidate for the energy.



\section{Overview and Outlook}
\label{sum}

Perhaps the most basic motivation for the study of quasilocal horizons is the philosophy that black holes should be 
thought of a physical objects in a spacetime which may be identified and quantified by local measurements. This
contrasts with causally defined black holes and event horizons which are defined from the asymptotic causal structure; with such a
definition, it is hard to imagine how physical processes could possibly be dependent on the location of the event horizon. Any such 
interactions would have to be non-local, almost certainly non-causal, and so violate some of our most fundamental ideas about physics.
This is then a strong argument for the non-physicality of event horizons and a need for some replacement. 

Of course there is a potential flaw in the above argument. The basis motivating idea could be wrong. It is certainly possible that while event horizons may not be particularly physical there may also be no replacement for them. 
In this case black holes would be fundamentally different
from other objects in the universe such as stars, elephants, and toasters. 
They would be symptomatic of strong gravitational fields but
have no meaning independent of those fields. 
Then, interactions with black holes could only be understood in the context of the 
global evolution of spacetime and it might not make sense to discuss such things as the exchange of energy and 
angular momentum
between a black hole and its environment. 

This seems unlikely. Results such as black hole mechanics, energy exchange 
mechanisms (eg.\! the Penrose process), 
a well-developed perturbation theory of known solutions, and an equally well-developed
infrastructure for calculating the gravitational waves emitted by black holes interacting with their surroundings all suggest
that, at least in some regime, we should be able to view black holes as objects for physical study. 

Given that the association of trapped surfaces with black holes actually precedes that of event horizons, they provide
the obvious candidate for a definitive black hole signature. In particular, it is well known that their appearance implies both 
spacetime singularities and the existence of event horizons. 
Further they play a key role in the definition of apparent horizons,
which are essentially the boundary of the region of trapped surfaces at some ``instant" of time. 
Then, we have argued that one can 
quasilocally identify a black hole boundary by finding a three-surface 
that is marginally trapped (with $\tl = 0$ and $\tn < 0$) and 
bounds a region in which all points lie on a trapped surface. 

A chief concern with this definition is that such surfaces are not invariantly defined. We have seen 
that this freedom may be constrained though certainly not entirely eliminated. 
Thus, it appears that we must accept that quasilocal 
black hole boundaries are non-unique and so, in a sense, observer dependent.\footnote
{The precise sense is as follows. Consider the MOTS to be outermost trapped surfaces as located in some $\Sigma_t$ foliation
of spacetime --- that is numerical/vernacular ``apparent" horizons. 
Further consider that foliation to be hypersurfaces of ``constant time" as determined by a fleet of observers with clocks
that were synchronized on some initial hypersurface. Then, as we have seen, different fleets will observe different MTTs.} 
This is not necessarily a problem --- as long as all set of observers agree in their conclusions/predictions (appropriately 
transformed between reference frames) no physical principles will be violated. 

That said, if one is chiefly interested in using these tools to better understand physical processes that may be seen by far-away observers,
we need to be a bit more careful. Since we are dealing with marginally trapped surfaces, causal signals emitted directly 
from these horizons cannot escape from the hole but instead must head further into its depths (or at best remain on the horizon). 
Therefore, anything seen by those far-away observers must be a consequence of the environment interacting with the hole's external field 
rather than the horizon directly (this applies to event horizons as well and is a key motivating 
idea behind the membrane paradigm \cite{membrane}). Thus, to be useful in trying to better understand far-field observations, our
quasilocal horizons must act as proxies for that near field. For this purpose, their utility is in characterizing that near field by 
closely reflecting its evolution. 

With this in mind, consider a highly dynamical process in which multiple horizons develop 
along the lines of the example shown in Fig.~\ref{R3}. 
Then there is no reason to expect the behaviour of inner horizons, buried deep inside the hole,
to be especially useful in interpreting, say, the gravitational waves emitted by this process. 
In fact, in a sufficiently dramatic evolution 
one probably wouldn't expect even the outermost horizon to closely reflect the near-field geometry. 
Based on these observations it
seems likely one could only expect a direct correlation between horizon behaviours and far-field observations 
in certain regimes --- one of
which would probably 
be slowly evolving horizons. In these regimes, however, it should be possible to ask (and answer) many interesting 
physical questions. 
For example, one could study how the multipole moments of the horizon \cite{multipoles}
determine how it interacts with its environment and gains/loses energy and angular momentum. 

Physically these are the most obviously relevant problems, however there are also interesting questions of a more mathematical interest. For example, we have seen that  it is possible to have (spherically symmetric) spacetimes with complicated MTTs that weave backwards and forwards through time and so appear as multiple horizons on a 
given hypersurface. Further, numerical simulations \cite{numerExamples} often show multiple horizons at a given
``instant". Thus it is natural to speculate that these may actually correspond to (a smaller number of) MTTs weaving
backwards and forwards in time, or even just intersecting the foliation in non-trivial ways. Other interesting topics for
investigation would include a more detailed understanding of the non-uniqueness of MTTs and the possibility that 
the trapping boundary might actually coincide with the event horizon. Such a result would be especially interesting as
it would provide an alternative way of identifying ``event horizons" --- without reference to global causal structure.  

These and other questions about quasilocal horizons can be cleanly defined and are 
mathematically both interesting and yet relatively accessible. Thus it seems likely that the next few years
will see a rapid development in our understanding of quasilocal horizons and a concordant growth of applications of these
ideas to physical and numerical problems.  

\section{Acknowledgements}
This work was supported by NSERC and is loosely based 
on a talk given by the author 
at the $11^{\mbox{th}}$ Canadian Conference on General Relativity and Relativistics Astrophysics (which 
was held at the University of British Columbia in May 2005).

\end{document}